\documentclass{elsarticle}

\usepackage{xspace}
\usepackage{comment}
\usepackage{lineno,hyperref}
\usepackage{array}
\usepackage{booktabs}
\usepackage{tcolorbox}
\usepackage{mathtools}
\usepackage{multirow}
\usepackage{subcaption}
\usepackage{soul}
\usepackage{color}
\modulolinenumbers[5]

\newcommand{\tool}{CoRec\xspace}
\newcommand{\toolu}{CoRec$_u$\xspace}
\newcommand{\codefont}[1]{\footnotesize{\texttt{#1}}\normalsize}

\newcolumntype{L}[1]{>{\raggedright\let\newline\\\arraybackslash\hspace{0pt}}m{#1}}
\newcolumntype{C}[1]{>{\centering\let\newline\\\arraybackslash\hspace{0pt}}m{#1}}
\newcolumntype{R}[1]{>{\raggedleft\let\newline\\\arraybackslash\hspace{0pt}}m{#1}}

\newcommand{\commentout} [1]{}
\newcommand{\red}[1]{{\textcolor{black}{#1}}}
\begin{document}

\begin{frontmatter}

\title{Investigating and Recommending Co-Changed Entities for JavaScript Programs}

\author[1]{Zijian Jiang}
\ead{wz649588@vt.edu}
\author[2]{Hao Zhong}
\ead{zhonghao@sjtu.edu.cn}
\author[1]{Na Meng\corref{cor1}}
\ead{nm8247@vt.edu}
\cortext[cor1]{Corresponding author}
\address[1]{Virginia Polytechnic Institute and State University, Blacksburg VA {\rm 24060}, USA}
\address[2]{Shanghai Jiao Tong University, Shanghai {\rm 200240}, China}

\begin{abstract}
JavaScript (JS) is one of the most popular programming languages due to its flexibility and versatility, but \red{maintaining} JS code is tedious and error-prone.
In our research, we conducted an empirical study to characterize the relationship between co-changed software entities (e.g., functions and variables), and built a machine learning (ML)-based approach to recommend additional entity to edit given developers' code changes. Specifically, we first crawled 14,747 commits in 10 open-source projects; for each commit, we created one or more change dependency graphs (CDGs) to model the referencer-referencee relationship between co-changed entities. Next, we extracted the common subgraphs between CDGs to locate recurring co-change patterns between entities. Finally, based on those patterns, we extracted code features from co-changed entities and trained an ML model that recommends entities-to-change given a program commit.


According to our empirical investigation, 
(1) three recurring patterns commonly exist in all projects; (2) 80\%--90\% of co-changed function pairs either invoke the same function(s), access the same variable(s), or contain similar statement(s); \red{(3)} our ML-based approach \tool recommended entity changes with high accuracy \red{(73\%--78\%)}. \red{\tool complements prior work because it suggests changes based on program syntax, textual similarity, as well as software history; it achieved higher accuracy than two existing tools in our evaluation. }




\end{abstract}


\end{frontmatter}


\section{Introduction}
\label{sec:intro}

JavaScript (JS) has become one of the most popular programming languages because it is lightweight, flexible, and powerful~\cite{js-popular}. Developers use JS to build web pages and games.
\red{JS has many new traits (1) it is dynamic and weakly typed; (2) it has first-class functions; (3) it is a class-free, object-oriented programming language that uses prototypal inheritance instead of classical inheritance; and (4) objects in JS inherit properties from other objects directly and all these inherited properties can be changed at runtime. All above-mentioned traits make JS unique and powerful; they also make JS programs very challenging to maintain and reason about~\cite{Amin2013,Saboury2017,Ferguson2019}.
}

\red{To reduce the cost of maintaining software,
researchers proposed approaches that recommend code co-changes~\cite{Zimmermann:2004,Rolfsnes:2018,Wang2018CMSuggester,Jiang2020}. For instance, Zimmermann et al.~\cite{Zimmermann:2004} and Rolfsnes et al.~\cite{Rolfsnes:2018} mined co-change patterns of program entities from software version history and suggested co-changes accordingly. Wang et al.~\cite{Wang2018CMSuggester,Jiang2020} studied the co-change patterns of Java program entities and built CMSuggester to suggest changes accordingly for any given program commit.
However, existing tools do not characterize any co-change patterns between JS software entities, neither do they recommend changes
by considering the unique language features of JS or the mined co-changed patterns from JS programs (see Section~\ref{sec:pre:mining}) for detailed discussions).}



\red{To overcome the limitations of the prior approaches,}
in this paper,
we first conducted a study on 14,747 program commits from 10 open-source JS projects to investigate (1) what software entities are usually edited together, and (2) how those simultaneously edited entities are related. \red{Based on this characterization study for co-change patterns, we further developed a learning-based approach \tool to recommend changes given a program commit.}

\red{Specifically in our study, for any program commit, we constructed and compared Abstract Syntax Trees (ASTs) for each edited JS file to identify all edited entities} (e.g., {Deleted Classes (DC)}, {Changed Functions (CF)}, and {Added Variables (AV)}). Next, we created \red{change dependency graphs (CDGs)} for each commit by treating edited entities as nodes and linking entities that have referencer-referencee relations. Afterwards, \red{we extracted common subgraphs between CDGs} and regarded those common subgraphs as recurring change patterns. In our study, we explored the following research question:

\begin{description}
\item \red{\textbf{RQ1: What are the frequent co-change patterns in JS programs?}}
\end{description}
\red{We automatically analyzed thousands of program commits from ten JS projects and revealed the recurring co-change patterns in each project. By manually inspecting 20  commits sampled for each of the 3 most popular patterns, we observed that 80\%--90\% of co-changed function pairs either invoke the same function(s), access the same variable(s), contain similar statement(s), or get frequently co-changed in version history. }

Besides the above findings, our study reveals three most popular change patterns:
\red{(i) one or more caller functions are changed together with one changed callee function that they commonly invoke; (ii) one or more functions are changed together to commonly invoke an added function; (iii) one or more functions are changed together to commonly access an added variable. The co-changed callers in each pattern may share commonality in terms of variable accesses, function invocations, code similarity, or evolution history.}
\commentout{\textbf{*CF$\xrightarrow{f}$CF} (i.e., one or more caller functions are changed together with one changed function that they commonly invoke), \textbf{*CF$\xrightarrow{f}$AF} (i.e., one or more functions are changed together to commonly invoke an added function), and \textbf{*CF$\xrightarrow{v}$AV} (i.e., one or more functions are changed together to commonly access an added variable).}


Based on the above-mentioned observations, we built a machine learning (ML)-based approach---\tool---to recommend functions for co-change. 
\red{Given the commits that contain matches for any of the above-mentioned co-change patterns,
\tool extracts 10 program features to characterize the co-changed function pairs, and uses those features to train an ML model.}
Afterwards,
 given a new program commit, the model predicts whether any unchanged function should be changed as well and recommends changes whenever possible. \red{With \tool, we investigated the following research question:}

\begin{description}
\item \red{\textbf{RQ2: How does \tool perform when suggesting co-changes based on the observed three most popular patterns?}}
\end{description}
\red{We applied \tool and two existing techniques (i.e., ROSE~\cite{Zimmermann:2004} and Transitive Associate Rules (TAR)~\cite{Islam:2018}) to the same evaluation datasets, and observed \tool to outperform both techniques by correctly suggesting many more changes. \tool's effectiveness varies significantly with the ML algorithm it adopts. \tool works better when it trains three separate ML models corresponding to the three patterns than training a unified ML model for all patterns.} Our results show that \tool can recommend co-change functions with 73--78\%
 accuracy; it significantly outperforms \red{two} baseline techniques that suggest co-changes purely based on software evolution.

We envision \tool to be used in the integrated development environments (IDE) for JS, code review systems, and version control systems. In this way, after developers make code changes or before they commit edits to software repositories, \tool can help  detect and fix incorrectly applied multi-entity edits.
In the sections below, we will first describe a motivating example (Section~\ref{sec:motivation}), and then introduce the \red{concepts used in} our research (Section~\ref{sec:concepts}). 
Next, we will present
the empirical study to characterize co-changes in JS programs (Section~\ref{sec:study}). Afterwards, we will explain our change recommendation approach \tool (Section~\ref{sec:approach})  and expound on the evaluation results (Section~\ref{sec:eval}). 

\section{A Motivating Example}
\label{sec:motivation}

\red{The prior work~\cite{Fry2010,Nguyen2010:fixwizard,Yin:2011,Park2012:supplementary} shows that developers may commit errors of omission (i.e., forgetting to apply edits completely) when they have to edit multiple program locations simultaneously in one maintenance task (i.e., bug fixing, code improvement, or feature addition). For instance, Fry et al.~\cite{Fry2010} reported that developers are over five times more precise at locating errors of commission than errors of omission. Yin et al.~\cite{Yin:2011} and Park et al.~\cite{Park2012:supplementary} separately showed that developers introduced new bugs when applying patches to fix existing bugs. In particular, Park et al.~inspected the supplementary bug fixes following the initial bug-fixing trials, and summarized nine major reasons to explain why the initial fixes were incorrect. Two of the nine reasons were about the incomplete program edits applied by developers.
}

\begin{figure}[h]
\centering
\vspace{0em}
	\includegraphics[width=\linewidth]{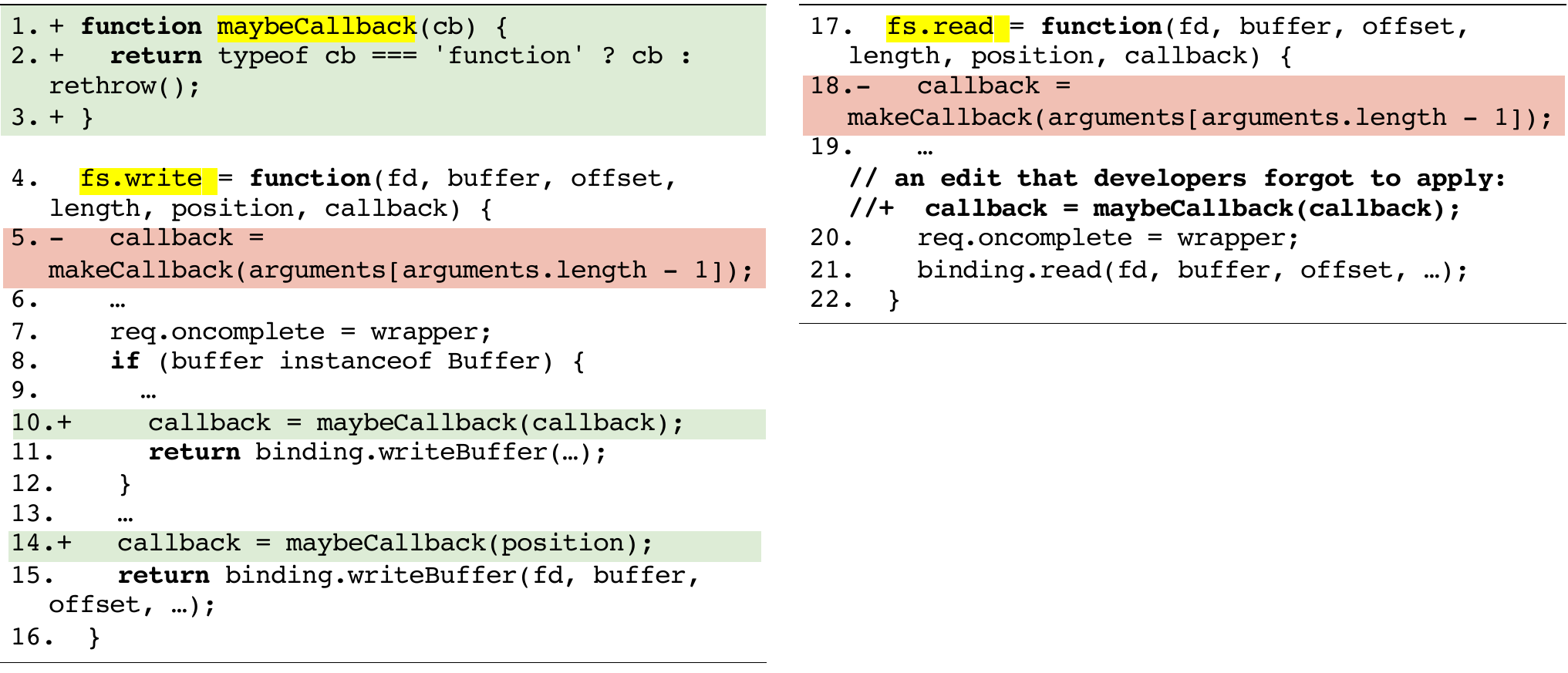}
	\caption{A program commit should add one function and change eight functions to invoke the newly added one.
	However, developers forgot to change one of the eight functions---\codefont{fs.read(...)}~\cite{node_21b0a27}.}	
	\label{fig:example}
	\vspace{0em}
	\end{figure}

\red{To help developers apply JS edits completely and avoid errors of omission, we designed and implemented a novel change recommendation approach---\tool. This section overviews our approach with a running example, which is extracted from a program commit to Node.js---an open-source server-side JS runtime environment~\cite{node}. Figure~\ref{fig:example} shows a simplified version of the exemplar program commit~\cite{node_21b0a27}}. In this revision, developers added a function \codefont{maybeCallback(...)} to check whether the pass-in parameter \codefont{cb} is a function, and modified seven functions in distinct ways to invoke the added function(e.g., changing \codefont{fs.write(...)} on line 10 and line 14). \red{The seven functions include: \codefont{fs.rmdir(...)}, \codefont{fs.appendFile(...)}, \codefont{fs.truncate(...)}, \codefont{fs.write(...)}, \codefont{fs.readFile(...)}, \codefont{fs.writeFile(...)}, and \codefont{fs.writeAll (...)}~\cite{node_21b0a27}. However, developers forgot to change an eighth function---\codefont{fs.read(...)}---to also invoke the added function (see line 19 in Figure~\ref{fig:example}). }

\red{\tool reveals the missing change with the following steps. \tool first trains an ML model with the program co-changes extracted from Node.js software version history. Then given the exemplar commit}, based on the added function \codefont{maybeCallback(...)} and each changed function (e.g., \codefont{fs.write(...)}), \tool extracts any commonality between the changed function and any unchanged one. \red{For each function pair, \tool applies its ML model to the extracted commonality features} and predicts whether the function pair should be changed together. Because \codefont{fs.write(...)} and \codefont{fs.read(...)}
\begin{itemize}
\item commonly access one variable \codefont{binding},
\item commonly invoke two functions: \codefont{makeCallback(...)} and \codefont{wrapper(...)},
\item declare the same parameters in sequence,
\item have token-level similarity as 41\%, and
\item have statement-level similarity as 42\%,
\end{itemize}
the pre-trained ML model inside \tool considers the two functions to share sufficient commonality and thus recommends developers to also change \codefont{fs.read(...)} to invoke \codefont{maybeCallback(...)}. In this way, \tool can
suggest entities for change, which edits developers may otherwise miss.

\section{\red{Terms and Definitions}}
\label{sec:concepts}

This section first introduces concepts relevant to JS programming, and then describes the terminology used in our research.

\textbf{ES6 and ES5.}
\red{ECMA Script is the standardized name for JavaScript~\cite{es6}. ES6 (or ECMAScript2015)
	is a major enhancement to ES5, and adds many features intended to make large-scale software development easier. ES5 is fully supported in all modern browsers, and
	major web browsers support some features of ES6. Our research is applicable to both ES5 and ES6 programs.}

\begin{figure}
	\includegraphics[width=\linewidth]{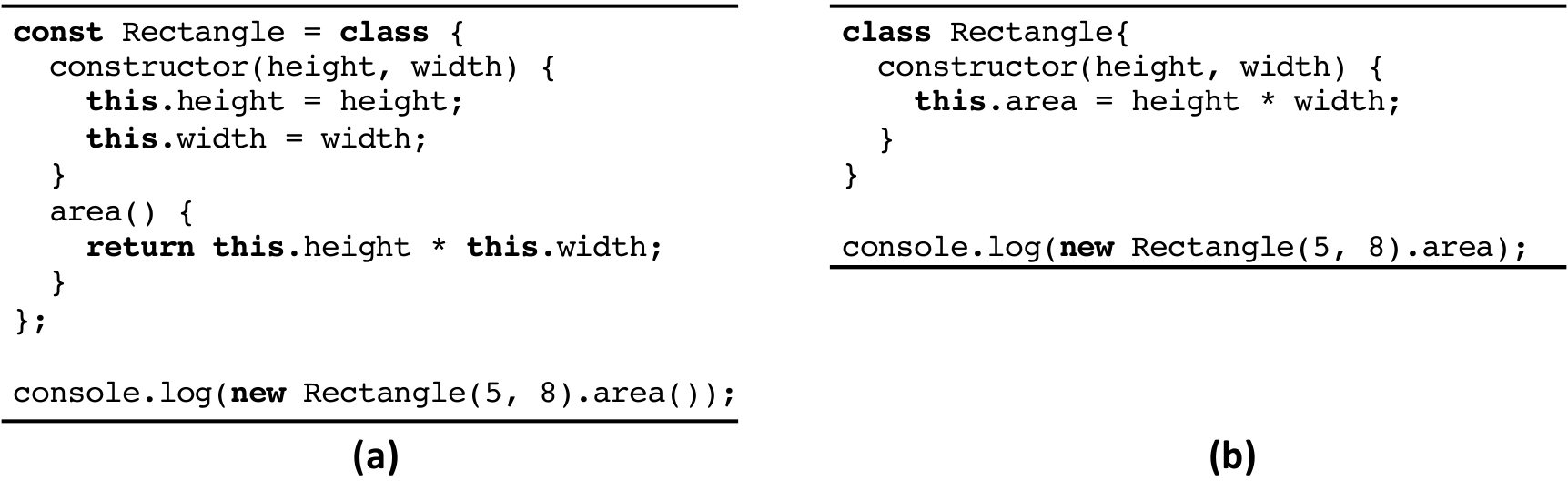}
	\caption{A JS class can be defined with an expression (see (a)) or a declaration (see (b)). }
	\label{fig:js-class}
\end{figure}

\textbf{Software Entity.}
\red{We use \emph{software entity} to refer to any defined JS \textbf{class}, \textbf{function}, \textbf{variable}, or any \textbf{independent statement block} that is not contained by the definition of classes, functions, or variables. When developers write JS code, they can define each type of entities in multiple alternative ways.}
 For instance, a class can be defined with a class expression (see Figure~\ref{fig:js-class} (a)) or class declaration (see Figure~\ref{fig:js-class} (b)). Similarly, a function can be defined with a function expression or function declaration. A variable can be defined with a variable declaration statement; the statement can either use keyword \codefont{const} to declare a constant variable, or use \codefont{let} or \codefont{var} to declare a non-constant variable. 

\textbf{Edited Entity.}
When maintaining JS software, developers may add, delete, or change one or more entities. Therefore, as with prior work~\cite{Ren:2004}, we defined a set of \textbf{\emph{edited entities}} to describe the possible entity-level edits, including \textit{Added Class} (\textbf{AC}), \textit{Deleted Class} (\textbf{DC}), \textit{Added Function} (\textbf{AF}), \textit{Deleted Function} (\textbf{DF}), \textit{Changed Function} (\textbf{CF}), \textit{Added Variable} (\textbf{AV}), \textit{Deleted Variable} (\textbf{DV}), \textit{Changed Variable} (\textbf{CV}), \textit{Added Statement Block} (\textbf{AB}), \textit{Deleted Statement Block} (\textbf{DB}), and \textit{Changed Statement Block} (\textbf{CB}). For example, if a new class is declared to have a constructor and some other methods, we consider the revision to have one AC, multiple AFs, and one or more AV (depending on how many fields are defined in the constructor).

\textbf{Multi-Entity Edit and CDG.}
As with prior work~\cite{yewang2018}, we use \textbf{\emph{multi-entity edit}} to refer to any commit that has two or more \textbf{\emph{edited entities}}. We use \textbf{\emph{change dependency graph (CDG)}} to visualize the the relationship between co-changed entities in a commit. Specifically, each CDG has at least two nodes and one edge. Each node represents an edited entity, and each edge represents the referencer-referencee relationship between entities (e.g., a function calls another function). Namely, if an edited entity $E_1$ refers to another edited entity $E_2$, we say $E_1$ depends on $E_2$. \red{A related CDG is constructed to connect the two entities with a directed edge pointing to $E_2$---the entity being depended upon (i.e. $E_1\rightarrow E_2$).}
For each program commit, we may create zero, one, or multiple CDGs.

\section{Characterization Study}
\label{sec:study}
\begin{figure}
	\centering
	\includegraphics[width=.7\linewidth]{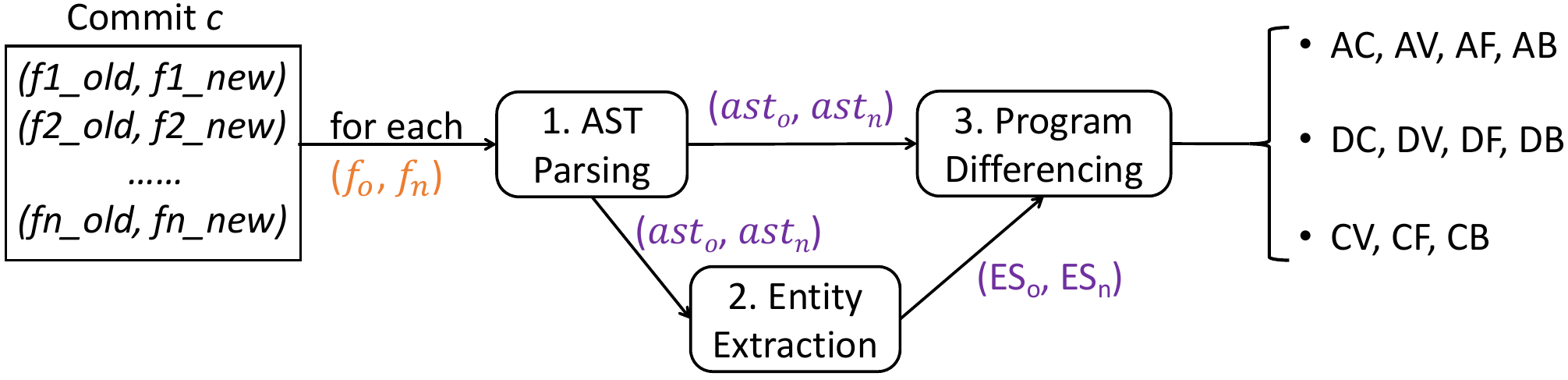}
	\caption{The procedure to extract changed entities given a commit.}
	\label{fig:entity-extractor}
\end{figure}
This section introduces our study methodology (Section~\ref{sec:method}) and explains the empirical findings (Section\red{~\ref{sec:empirical}}). The purpose of this characterization study is to identify \textbf{\emph{recurring change pattern (RCP)}} of JS programs. An RCP is a CDG subgraph that is commonly shared by the CDGs from at least two distinct commits. RCPs define different types of edits, and serve as the templates of co-change rules. Our approach in Section~\ref{sec:approach} mines concrete co-change rules for the most common RCPs.

\subsection{Study Methodology}
\label{sec:method}
We implemented a tool to automate the analysis. Given a set of program commits in JS repositories, our tool first characterizes each commit by extracting the edited entities (Section~\ref{sec:extract}) and constructing CDG(s) (Section~\ref{sec:cdg}). Next, it compares CDGs across commits to identify RCPs (Section~\ref{sec:rcp}).

\subsubsection{Extraction of Edited Entities}
\label{sec:extract}

\red{As shown in Figure~\ref{fig:entity-extractor}, we took three steps to extract any edited entities for each commit.}

\paragraph*{\textbf{\red{Step 1: AST Parsing}}}
Given a program commit $c$, this step first locates the old and new versions of each edited JS file. For every edited file $(f_o, f_n)$, this step adopts Esprima~\cite{esprima} and typed-ast-util~\cite{typed-ast} to generate Abstract Syntax Trees (ASTs): $(ast_o, ast_n)$.
Esprima is a high performance, standard-compliant JavaScript parser that supports the syntax of both ES5 and ES6;  
however, it cannot infer the static type binding information of any referenced class, function, or variable.
 Meanwhile, given JS files and the project's \codefont{package.json} file, typed-ast-util produces ASTs annotated with structured representations of \red{TypeScript} types, which information can facilitate us to precisely identify the referencer-referencee relationship between edited entities. 
We decided to use both tools for two reasons. First, when a project has \codefont{package.json} file, we rely on Esprima to identify the code range and token information for each parsed AST node, and rely on typed-ast-util to attach relevant type information to those nodes. Second, if a project has no \codefont{package.json} file, Esprima is still used to generate ASTs but we defined a heuristic approach (to be discussed later in Section~\ref{sec:cdg}) to identify the referencer-referencee relationship between entities with best efforts.

\begin{figure}
	\centering
	\includegraphics[width=1\linewidth]{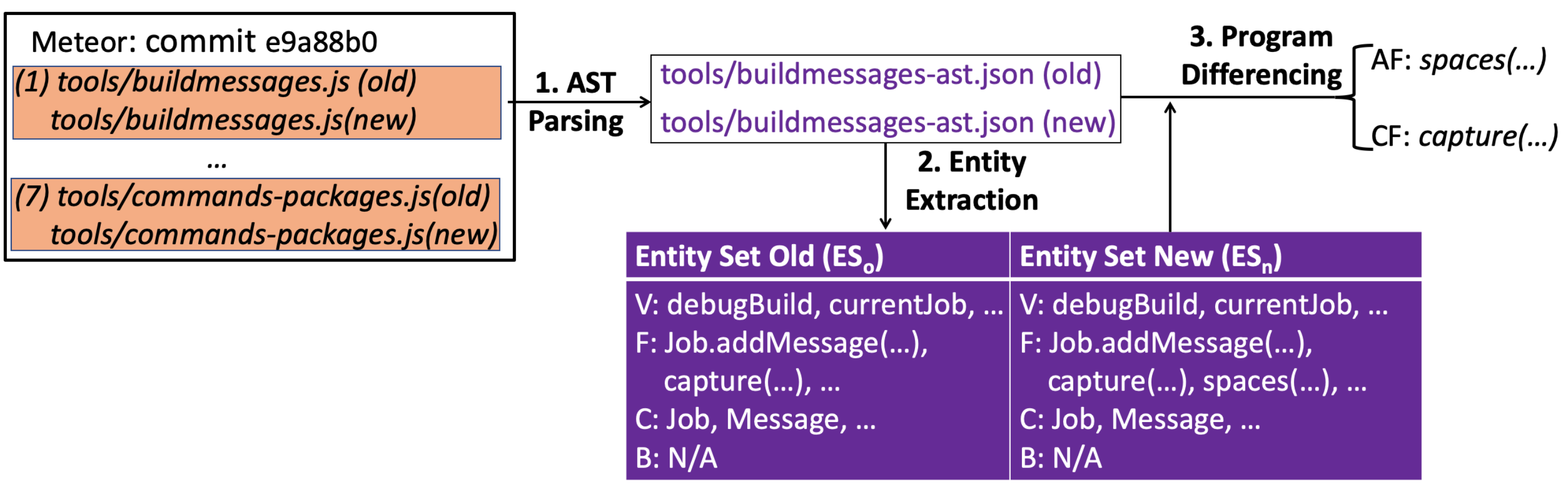}
	\caption{Extracting edited entities from a program commit of Meteor~\cite{e9a88b0}.}
	\label{fig:CEExample}
\end{figure}

\red{To facilitate our discussion, we introduce a
working example from a program revision~\cite{e9a88b0} of Meteor~\cite{meteor}. As shown in Figure~\ref{fig:CEExample}, the program revision changes seven JS files. In this step, \tool creates a pair of ASTs for each edited file and stores the ASTs into JSON files for later processing (e.g., \codefont{tools/buildmessages-ast.json (old)} and \codefont{tools/buildmessages-ast.json (new)}).
}

\begin{figure}
	\centering
	\includegraphics[width=.9\linewidth]{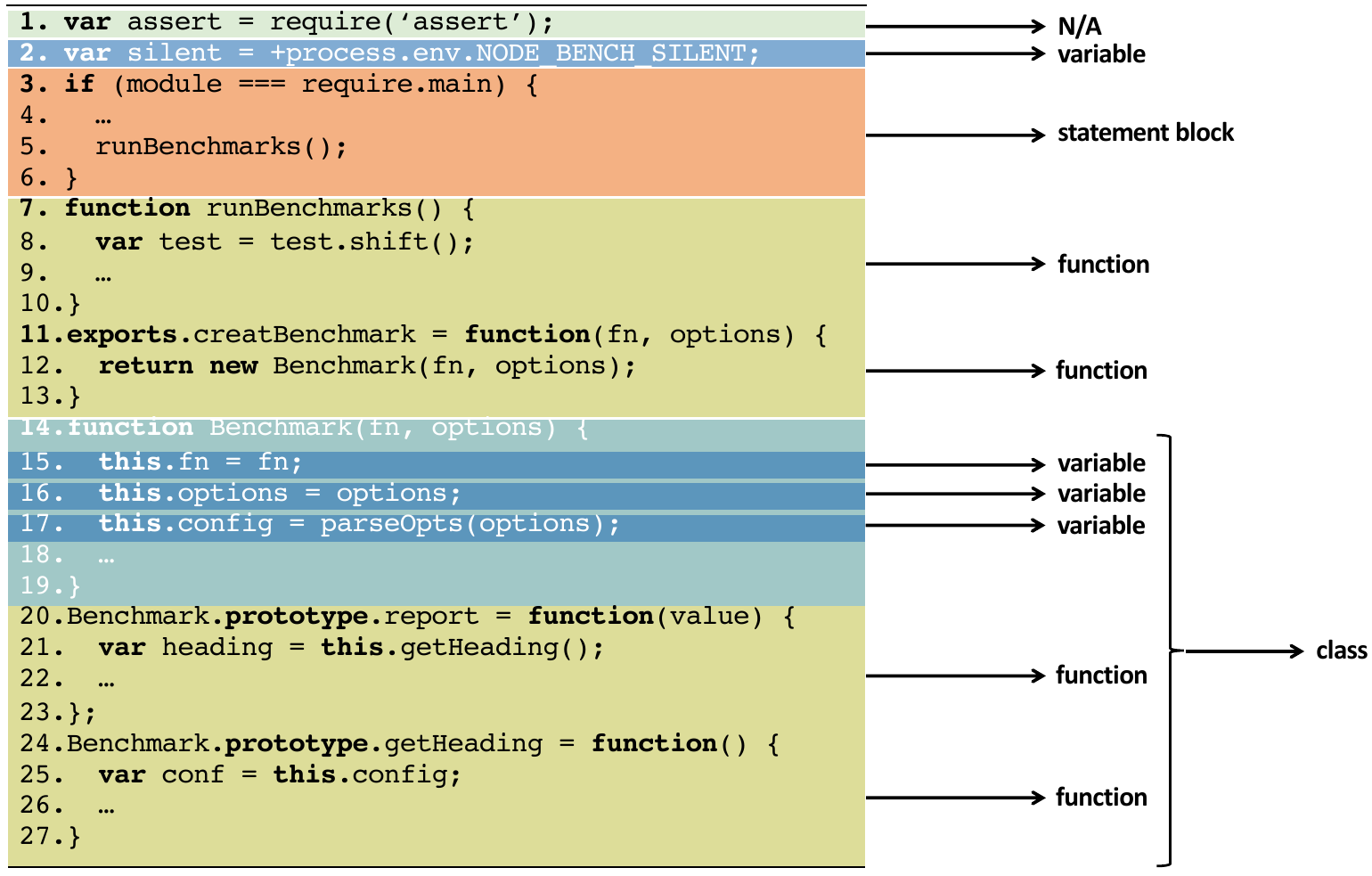}
	\caption{Code snippets from the file {\tt benchmark.common.js} of 	
	 Node.js in revision 00a1d36~\cite{node_21b0a27}, whose related entity types are shown on the right.}
	\label{fig:entity-examples}
\end{figure}

\paragraph*{\textbf{\red{Step 2: Entity Extraction}}}
From each pair of ASTs $(ast_o, ast_n)$ (i.e., JSON files), this step extracts the entity sets $(ES_o, ES_n)$. \red{In the example shown in Figure~\ref{fig:CEExample}, $ES_o$ lists all entities from the old JS file, and $ES_n$ corresponds to the new file. We defined four kinds of entities to extract:  variables (V), functions (F), classes (C), and statement blocks (B). A major technical challenge here is \emph{how to extract entities precisely and consistently}. Because JS programming supports diverse ways of defining entities and the JS syntax is very flexible, we cannot simply check AST node types of statements to recognize entity definitions.}
For instance, a variable declaration statement can be interpreted as a variable-typed entity or a statement block, depending on the program context. To eliminate ambiguity and avoid any confusion between differently typed entities, we classify and extract entities in the following way:

\begin{itemize}
\item A code block is treated as a function definition if it satisfies either of the following two requirements. First, the AST node type is ``\codefont{FunctionDeclaration}'' \red{(e.g., \codefont{runBenchmarks()} on line 7 in Figure~\ref{fig:entity-examples})} or ``\codefont{MethodDefinition}''. Second, (1) the block is either a ``\codefont{VariableDeclaration}'' statement (e.g., \codefont{const getRectArea = function(...)\{...\};}) or an``\codefont{Assignment}'' expression \red{(see line 11 and line 20 of Figure~\ref{fig:entity-examples}}); 
and (2) the right-side operand is either ``\codefont{FunctionExpression}'', or ``\codefont{CallExpression}'' that outputs another function as return value of the called function. In particular, if any defined function has its prototype property explicitly referenced \red{(e.g., \codefont{Benchmark.prototype} on lines 20 and 24)}
or is used as a constructor to create any object \red{(e.g., line 12)},
we reclassify the function definition as a class definition, because the function usage is more like the usage of a class.

\item A code block is considered to be a class definition if it meets either of the following two criteria. First, the block uses keyword \codefont{class}. Second, the block defines a function, while the codebase either references the function's prototype \red{(e.g., \codefont{Benchmark.prototype} on lines 20 and 24 in Figure~\ref{fig:entity-examples})} or uses the function as a constructor to create any object \red{(see line 12)}.

\item A code block is treated as a variable declaration if (1) it
is either a ``\codefont{VariableDeclaration}'' statement \red{(e.g., \codefont{var silent = ...} on line 2 in Figure~\ref{fig:entity-examples}}) or an ``\codefont{Assignment}'' expression, (2) it does not define a function or class, (3) it does not belong to the definition of any function but may belong to a constructor \red{(see lines 15-17)}, and (4) it does not declare a required module \red{(see  line 1)}.
Particularly, when a variable declaration is an assignment inside a class constructor \red{(e.g., lines 15-17)},
it is similar to the field declaration in Java.

\item A code block is treated as a statement block if (1) it purely contains statements, (2) it does not define any class, function, or variable, and (3) it does not belong to the definition of any class or function. \red{For example, lines 3-6 in Figure~\ref{fig:entity-examples} are classified as a statement block.}
\end{itemize}

\paragraph*{\textbf{\red{Step 3: Program Differencing}}}
To identify any edited entity between $ES_o$ and $ES_n$, we first matched the definitions of functions, variables, and classes across entity sets based on their signatures. If any of these entities (e.g., a function definition) solely exists in $ES_o$, an entity-level deletion (e.g., DF) is inferred; if an entity (e.g., a variable definition) solely exists in $ES_n$, an entity-level insertion (e.g., AV) is inferred. Next, for each pair of matched entities, we further exploited a fine-grained AST differencing tool---GumTree~\cite{Falleri2014:GumTree}---to identify expression-level and statement-level edits. If any edit is reported, we inferred an entity-level change \red{(e.g., CF shown in Figure~\ref{fig:CEExample})}.
Additionally, we matched statement blocks across entity sets based on their string similarities. Namely, if a statement block $b_1 \in ES_o$ has the longest common subsequence with a block $b_2 \in ES_n$ and the string similarity is above 50\%, we considered the two blocks to match. Furthermore, if the similarity between two matched blocks is not 100\%, we inferred a block-level change CB.

\subsubsection{CDG Construction}
\label{sec:cdg}
For each program commit, we built CDGs by representing the edited entities as nodes, and by connecting edited entities with directed edges if they have either of the following two kinds of relationship:

\begin{itemize}
	\item \textbf{Access.} If an entity $E_1$ accesses another entity $E_2$ (i.e., by reading/writing a variable, invoking a function, or using a class), we consider $E_1$ to be dependent on $E_2$.
	\item \textbf{Containment.} If the code region of $E_1$ is fully covered by that of $E_2$, we consider $E_1$ to be dependent on $E_2$.
\end{itemize}

The technical challenge here is how to identify the relationship between edited entities.
We relied on ESprima's outputs to compare code regions between edited entities in order to reveal the containment relations. Additionally, when \codefont{package.json} file is available, we leveraged the type binding information inferred by typed-ast-util to identify the access relationship. For instance, if there is a function call \codefont{bar()} inside an entity $E_1$ while \codefont{bar()} is defined by a JS module \codefont{f2}, then typed-ast-util can resolve the fully qualified name of the callee function as \codefont{f2.bar()}. Such resolution enables us to effectively link edited entities no matter whether they are defined in the same module (i.e., JS file) or not.

Since some JS projects have no \codefont{package.json} file, we could not adopt typed-ast-util to resolve bindings in such scenarios. Therefore, we also built a simpler but more applicable approach to automatically speculate the type binding information of accessed entities as much as possible. Specifically, suppose that file \codefont{f1} defines $E_1$ to access $E_2$. To resolve $E_2$ and link $E_1$ with $E_2$'s definition, this intuitive approach first scans all entities defined in \codefont{f1} to see whether there is any $E_2$ definition locally. If not, this approach further examines all \codefont{require} and \codefont{import} statements in \codefont{f1}, and checks whether any required or imported module defines a corresponding entity with $E_2$'s name; if so, this approach links $E_1$ with the retrieved $E_2$'s definition.

Compared with typed-ast-util, our approach is less precise because it cannot infer the return type of any invoked function. For instance, if we have \codefont{const foo = bar()} where \codefont{bar()} returns a function, our approach simply assumes \codefont{foo} to be a variable instead of a function. Consequently, our approach is unable to link \codefont{foo}'s definition with any of its invocations.
\red{Based on our experience of applying both typed-ast-util and the heuristic method to the same codebase (i.e., nine open-source projects), the differences between these two methods' results account for no more than 5\% of all edited entities.
It means that our heuristic method is still very precise even though no \codefont{package.json} file is available}.

\begin{figure}
	\includegraphics[width=\linewidth]{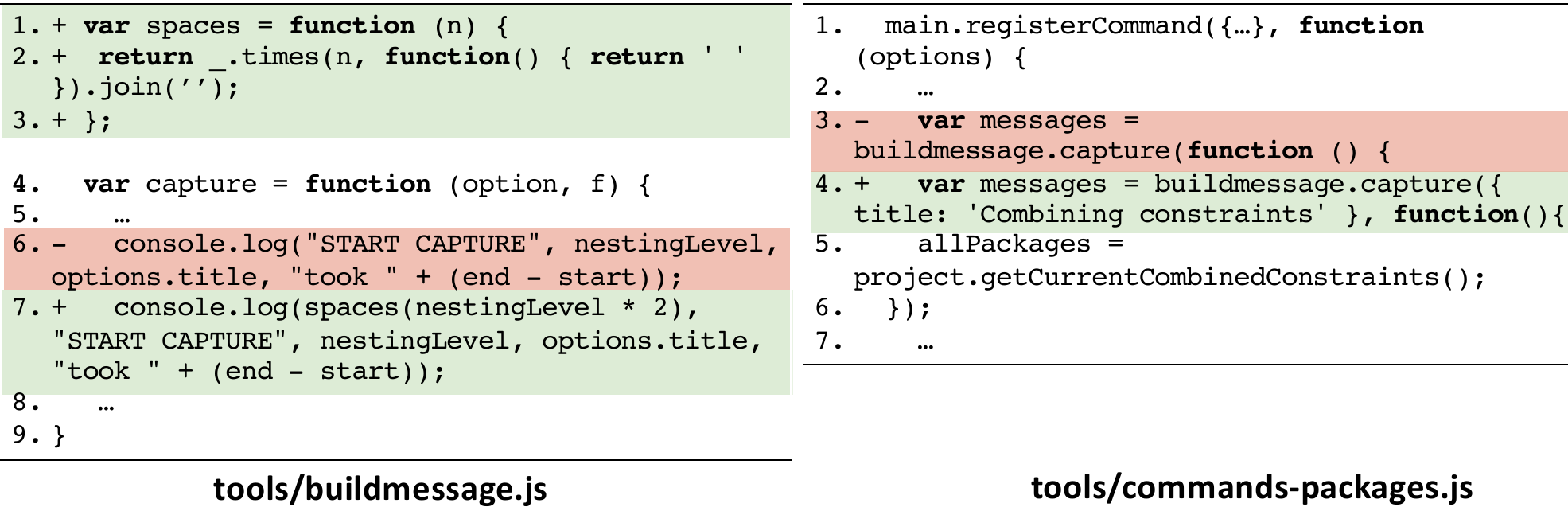}
	\caption{A simplified program commit that adds a function {\tt spaces(...)}, changes a function {\tt capture(...)}, and changes a statement block~\cite{e9a88b0}}
	\label{fig:cdg-example-code}
\end{figure}

\begin{figure}
	\centering
	\includegraphics[width=0.95\linewidth]{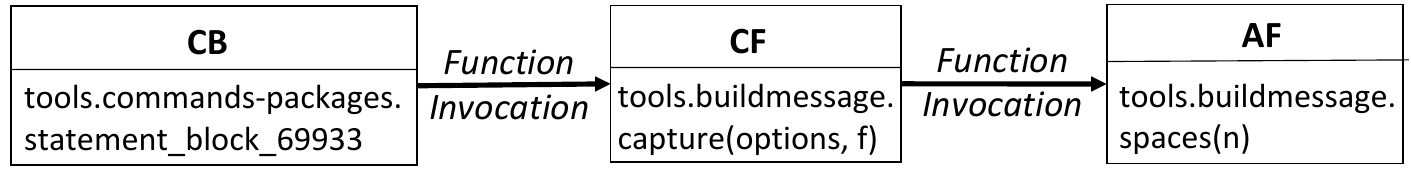}
	\caption{The CDG corresponding to the program commit shown in Figure~\ref{fig:cdg-example-code}}
	\label{fig:code-example-graph}
\end{figure}

\red{Figures~\ref{fig:cdg-example-code} and \ref{fig:code-example-graph} separately present the code changes and CDG related to \codefont{tools/buildmessage.js}, an edited file mentioned in Figure~\ref{fig:CEExample}.}
According to Figure~\ref{fig:cdg-example-code}, the program commit modifies file \codefont{tools/buildmessage.js} by defining a new function \codefont{spaces(...)} and updating an existing function \codefont{capture(...)} to invoke the new function. It also changes file \codefont{tools/commands-package.js} by updating the function invocation of \codefont{capture(...)} inside a statement block (i.e., \codefont{main.registerCommand(...);}). Given the old and new versions of both edited JS files, our approach can construct the CDG shown in Figure~\ref{fig:code-example-graph}. In this CDG, each directed edge starts from a dependent entity $E_1$, and points to the entity on which $E_1$ depends. Each involved function, variable, or class has its fully qualified name included in the CDG for clarity. As statement blocks have no fully qualified names, we created a unique identifier for each block with (1) the module name (e.g., \codefont{tools.commands-packages}) and (2) index of the block's first character in that module (e.g., \codefont{69933}).

\subsubsection{Extraction of Recurring Change Patterns (RCP)}\label{sec:rcp}
As with prior work~\cite{yewang2018}, we extracted RCPs by comparing CDGs across program commits. Intuitively, given a CDG $g_1$ from commit $c_1$ and the CDG $g_2$ from commit $c_2$, we matched nodes based on their edit-entity labels (e.g., AF) while ignoring the concrete code details (e.g., \codefont{tools.buildmessage.spaces(n)} in Figure~\ref{fig:code-example-graph}). We then established edge matches based on those node matches. Namely, two edges are matched only if they have matching starting nodes and matching ending nodes. Next, based on all established matches, we identified the largest common subgraph between $g_1$ and $g_2$ using the off-the-shelf subgraph isomorphism algorithm VF2~\cite{Cordella2004:VF2}. Such largest common subgraphs are considered as RCPs because they commonly exist in CDGs of different commits.

\begin{table}
	\centering
	\caption{Subject projects}
	\label{tab:projects}
	\scriptsize	
	\begin{tabular}{L{1.5cm} L{5.1cm} R{.7cm} R{1.2cm} R{1.5cm}}
		\toprule
		\textbf{Project} &\textbf{Description} & \textbf{\# of KLOC} & \textbf{\# of Commits} & \textbf{\# of Edited Entities}\\
		\toprule
		Node.js & Node.js~\cite{node} is a cross-platform JS runtime environment. It executes JS code outside of a browser. & 1,755 & 2,701 & 11,287\\
		\midrule
		Meteor & Meteor~\cite{meteor} is an ultra-simple environment for building modern web applications. & 255 & 3,011 & 10,274\\
		\midrule
		Ghost & Ghost~\cite{ghost} is the most popular open-source and headless Node.js content management system (CMS) for professional publishing. & 115 & 1,263 & 5,142\\
		\midrule
	    Habitica & Habitica~\cite{habitica} is a habit building program that treats people's life like a Role Playing Game. & 129 & 1,858 & 6,116\\
		\midrule
		PDF.js & PDF.js~\cite{pdf} is a PDF viewer that is built with HTML5. & 104 & 1,754 & 4,255\\
		\midrule
		React & React~\cite{react} is a JS library for building user interfaces. & 286 & 1,050 & 4,415\\
		\midrule
	    Serverless & Serverless~\cite{serverless} is a framework used to build applications comprised of microservices that run in response to events. &63 & 1,110 & 3,846\\
		\midrule
		Webpack & Webpack~\cite{webpack} is a module bundler, which mainly bundles JS files for usage in a browser. assets.  & 37 & 1,099 & 3,699\\
		\midrule
		Storybook & Storybook~\cite{storybook} is a development environment for UI components.  & 43 & 528 & 2,277\\
		\midrule
		Electron & Electron~\cite{electron} is a framework that supports developers to write cross-platform desktop applications using JS, HTML, and CSS. & 35 & 673 & 1,898\\
		\bottomrule
	\end{tabular}
	\vspace{-0.5em}
\end{table}

\subsection{Empirical Findings}
\label{sec:empirical}
\commentout{We analyzed the bug-fixing commits of 10 open-source projects, as shown in Table~\ref{tab:projects}. We chose these projects because they are popularly used and from different application domains. We mainly focused on bug-fixing commits because based on our experience, developers are more likely to check in related changes for one software maintenance task (i.e., bug fixing) in each of such commits. To identify bug-fixing commits, we searched for commits whose commit messages contain any of the following keywords: ``bug'', ``fix'', ``error'', ``adjust'', and ``failure''.}
\red{To characterize JS code changes, we applied our study approach to a subset of available commits in 10 open-source projects, as shown in Table~\ref{tab:projects}. We chose these projects because (1) they are popularly used; (2) they are from different application domains; and (3) they contain a large number of available commits. For simplicity, to sample the commits that may fulfill independent maintenance tasks, we searched each software repository for commits whose messages contain any of the following keywords: ``bug'', ``fix'', ``error'', ``adjust'', and ``failure''.}

\red{Table~\ref{tab:projects} shows the statistics related to the sampled commits. In particular, column \textbf{\# of KLOC} presents the code size of each project (i.e., the number of kilo lines of code (KLOC)). Column \textbf{\# of Commits} reports the number of commits identified via our keyword-based search. Column \textbf{\# of Edited Entities} reports the number of edited entities extracted from those sampled commits.}
According to this table, the code size of projects varies significantly from 35 KLOC to 1755 KLOC. Among the 10 projects, 528--3,011 commits were sampled, and 1,898--11,287 edited entities were included for each project.
Within these projects, only Node.js has no \codefont{package.json} file, so we adopted our heuristic approach mentioned in Section~\ref{sec:cdg} to link edited entities. For the remaining nine projects, as they all have \codefont{package.json} files, we leveraged the type binding information inferred by typed-util-ast  to connect edited entities.

\begin{figure}
\centering
\includegraphics[width=0.85\linewidth]{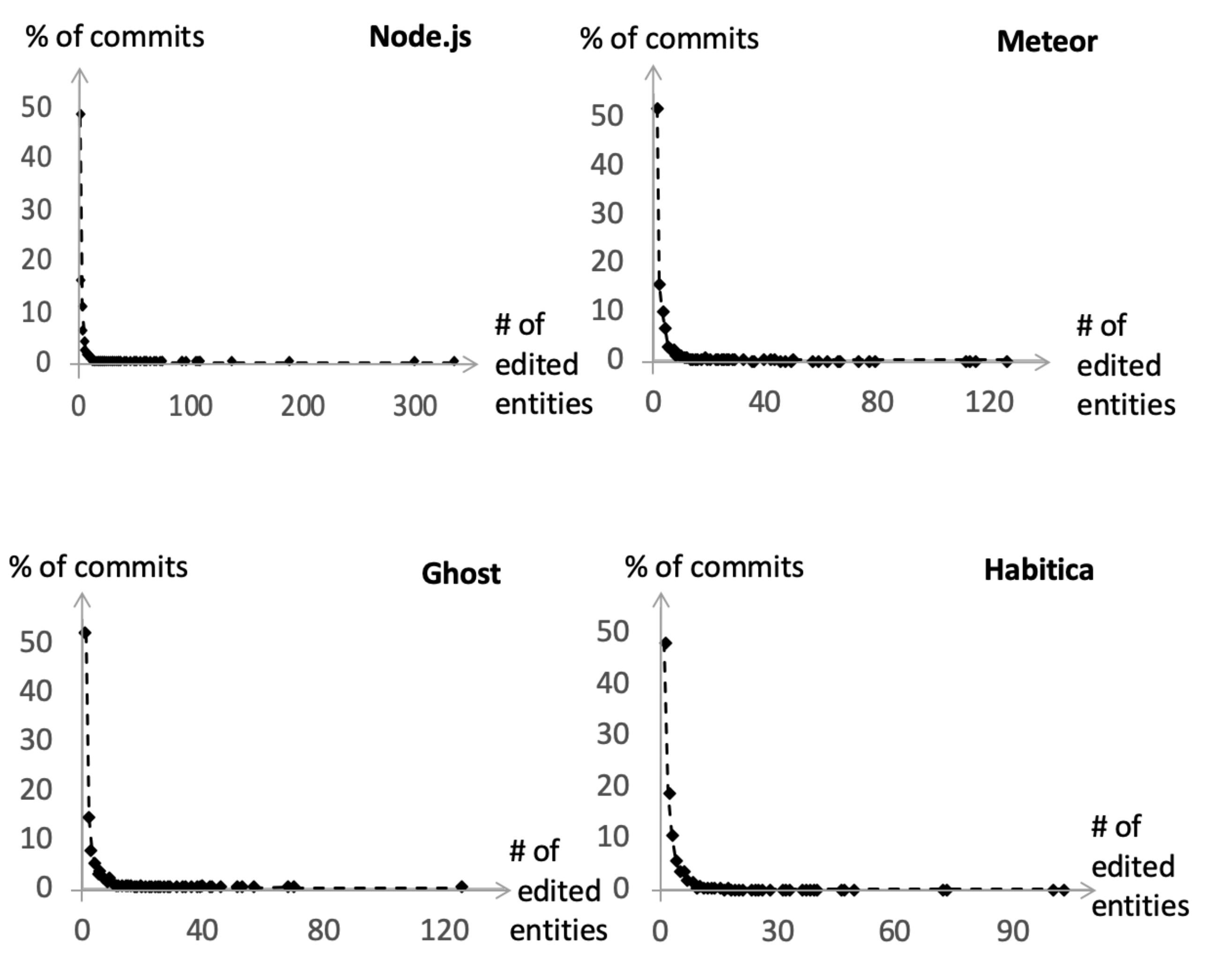}
\caption{Commit distributions based on the number of edited entities each of them contains}
\label{fig:distribute-by-entity}
\end{figure}

\subsubsection{Commit Distributions Based on The Number of Edited Entities}
\label{sec:distribution-by-entity}
We first clustered commits based on the number of edited entities they contain.
Because the commit distributions of different projects are very similar to each other, we present the distributions for four projects in Figure~\ref{fig:distribute-by-entity}. Among the 10 projects, 41\%--52\% of commits are multi-entity edits. Specifically, 15\%--19\% of commits involve two-entity edits, and 7\%--10\% of commits are three-entity edits.
The number of commits decreases as the number of edited entities increases.
The maximum number of edited entities appears in Node.js, where a single commit modifies 335 entities. We manually checked the commit on GitHub~\cite{e71e71b}, and found that  four JS files were added and three other JS files were changed to implement HTTP/2.

\red{We noticed that about half of sampled program revisions involve multi-entity edits. This observation implies the importance of co-change recommendation tools. When developers have to frequently edit multiple entities simultaneously to achieve a single maintenance goal, it is crucially important to provide automatic tool support that can check for the completeness of code changes and suggest any missing change when possible. In order to build such tools, we decided to further explore relations between co-changed entities (see Section~\ref{sec:cdg-distribution}).}

\begin{tcolorbox}
	\textbf{Finding 1:} \emph{Among the 10 studied projects, 41--52\% of studied commits are multi-entity edits. It indicates the necessity of our research to characterize multi-entity edits and  to recommend changes accordingly. }
\end{tcolorbox}

\begin{table}
	\centering
	\caption{Multi-entity edits and created CDGs}
	\label{tab:CDGs}
	\scriptsize	
	\begin{tabular}{L{1.5cm} R{3cm} R{2.8cm} R{3cm}}
		\toprule
		\textbf{Project} & \textbf{\# of Multi-Entity Edits} & \textbf{\# of Multi-Entity Edits with CDG(s) Extracted} & \textbf{\% of Multi-Entity Edits with CDG(s) Extracted}\\
		\toprule
		Node.js & 1,401 & 785 & 56\%\\
		\midrule
		Metoer & 1,445 & 670 & 46\%\\
		\midrule
		Ghost & 604 & 356 & 59\%\\
		\midrule
	    Habitica & 962 & 345 & 36\%\\
		\midrule
		PDF.js & 711 & 372 & 52\%\\
		\midrule
		React & 538 & 320 & 60\%\\
		\midrule
	    Serverless & 480 & 171 & 36\%\\
		\midrule
		Webpack & 483 & 253 & 52\%\\
		\midrule
		Storybook  & 243 & 119 & 49\%\\
		\midrule
		Electron & 277 & 123 & 44\%\\
		\bottomrule
	\end{tabular}
	\vspace{-0.5em}
\end{table}

\subsubsection{Commit Distributions Based on The Number of CDGs}\label{sec:cdg-distribution}
\begin{figure}
\centering
\includegraphics[width=0.85\linewidth]{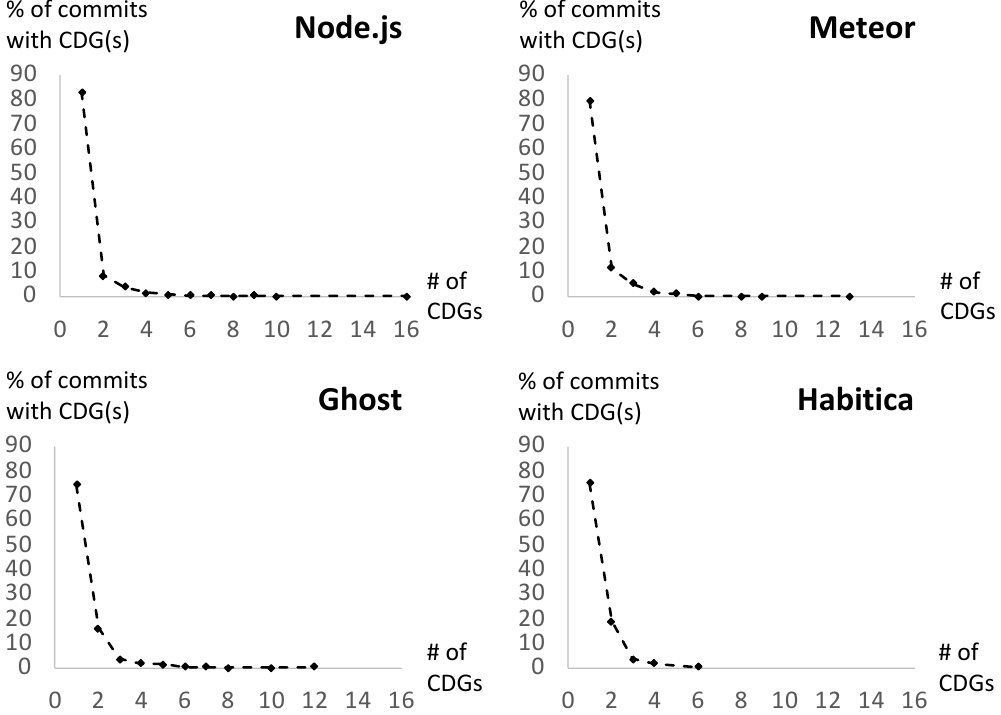}
\caption{The distributions of multi-entity edits based on the number of CDGs}
\label{fig:distribute-by-cdg}
\end{figure}

We further clustered multi-entity edits based on the number of CDGs constructed for each commit. As shown in Table~\ref{tab:CDGs}, our approach created CDGs for 36--60\% of the multi-entity edits in distinct projects. On average, 49\% of multi-entity edits contain at least one CDG. Due to the complexity and flexibility of the JS programming language, it is very challenging to statically infer all possible referencer-referencee relationship between JS entities. Therefore, the actual percentage of edits that contain related co-changed entities can be even higher than our measurement. Figure~\ref{fig:distribute-by-cdg} presents the distributions of multi-entity edits based on the number of CDGs extracted. Although this figure only presents the commit distributions for four projects: Node.js, Meteor, Ghost, and Habitica, we observed similar distributions in other projects as well.
As shown in this figure, the number of commits decreases significantly as the number of CDGs increases. Among all 10 projects, 73\%--81\% of commits contain single CDGs, 9\%--18\% of commits have two CDGs extracted, and 3\%--7\% of commits have three CDGs each. The commit with the largest number of CDGs constructed (i.e., 16) is the one with the maximum number of edited entities in Node.js as mentioned above in Section~\ref{sec:distribution-by-entity}.


\red{The high percentage of multi-entity edits with CDGs extracted (i.e., 49\%) implies that JS programmers usually change syntactically related entities simultaneously in program revisions. Such syntactic relevance between co-changed entities enlightened us to build tools that recommend co-changes by observing the syntactic dependences between changed and unchanged program entities. To concretize our approach design for co-change recommendations, we further explored the recurring syntactic relevance patterns between co-changed entities, i.e., RCPs (see Section~\ref{sec:identify-rcps}).}

\begin{tcolorbox}
	\textbf{Finding 2:} \emph{For 36--60\% of multi-entity edits in the studied projects, our approach created at least one CDG for each commit. It means that many simultaneously edited entities are syntactically relevant to each other.
}
\end{tcolorbox}

\begin{table}
	\centering
	\caption{Recurring change patterns and their matches}	
	\label{tab:patterns}
	\scriptsize
	\begin{tabular}{L{1.4cm} R{1.5cm} R{3cm} R{4cm}}
		\toprule
		\textbf{Projects} & \textbf{\# of RCPs} & \textbf{\# of Commits with RCP Matches} & \textbf{\# of Subgraphs Matching the RCPs}\\
		\toprule
		Node.js & 221 & 782 & 2,385\\
		\midrule
		Metoer & 200 & 658 & 1,719\\
		\midrule
		Ghost & 133 & 351 & 1,223\\
		\midrule
	    Habitica & 116 & 339 & 706\\
		\midrule
		PDF.js & 86 & 367 & 640\\
		\midrule
		React & 110 & 316 & 899\\
		\midrule
	    Serverless & 57 & 164 & 372\\
		\midrule
		Webpack & 80 & 243 & 583\\
		\midrule
		Storybook  & 42 & 113 & 337\\
		\midrule
		Electron & 35 & 117 & 228\\
		\bottomrule
	\end{tabular}
	\vspace{-0.5em}
\end{table}

\subsubsection{Identified RCPs}\label{sec:identify-rcps}
By comparing CDGs of distinct commits within the same project repository, we identified RCPs in all projects. As listed in Table~\ref{tab:patterns}, 35--221 RCPs are extracted from individual projects. In each project, there are 113--782 commits that contain matches for RCPs. In particular, each project has 228--2,385 subgraphs matching RCPs. By comparing this table with Table~\ref{tab:CDGs}, we found that 95\%--100\% of the commits with CDGs extracted have matches for RCPs. It means that if one or more CDGs can be built for a commit, the commit is very likely to share common subgraphs with some other commits. In other words, simultaneously edited entities are usually correlated with each other in a fixed number of ways. If we can characterize the frequently occurring relationship between co-changed entities, we may be able to leverage such characterization to predict co-changes or reveal missing changes.

\begin{figure}
\centering
\vspace{0em}
	\includegraphics[width=8cm]{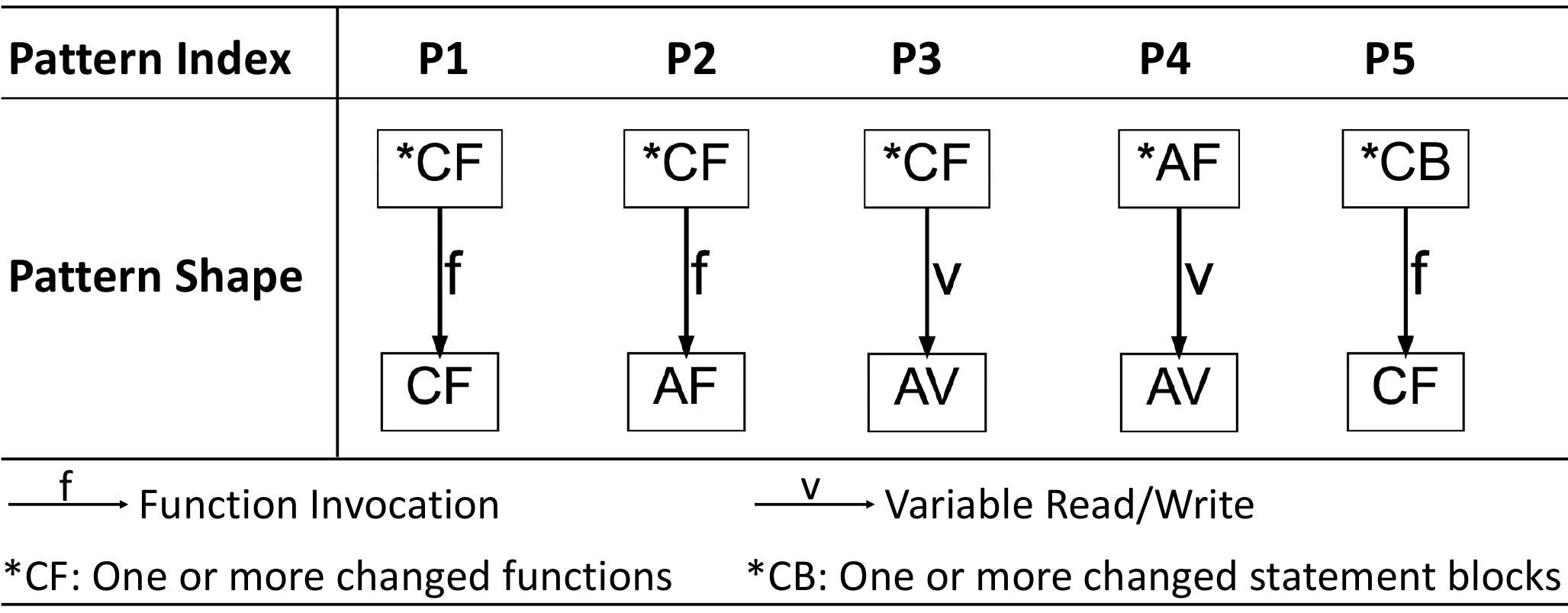}
	\caption{5 most popular recurring change patterns among the 10 projects}	
	\label{fig:popular}
	\vspace{0em}
\end{figure}

By accumulating the subgraph matches for RCPs across projects, we identified five most popular RCPs, as shown in Figure~\ref{fig:popular}.
Here, \textbf{P1} means that {when a callee function is changed, one or more of its caller functions are also changed.} \textbf{P2} means that when a new function is added, one or more existing functions are changed to invoke that new function. \textbf{P3} shows that {when a new variable is added, one or more existing functions are changed to read/write the new variable.} \textbf{P4} presents that {when a new variable is added, one or more new functions are added to read/write the new variable.} \textbf{P5} implies that {when a function is changed, one or more existing statement blocks invoking the function are also changed.}
Interestingly,
the top three patterns commonly exist in all 10 projects, while the other two patterns do not exist in some of the projects. The top three patterns all involve simultaneously changed functions.

\begin{tcolorbox}
	\textbf{Finding 3:} \emph{Among the commits with CDGs extracted, 95\%--100\% of commits have matches for mined RCPs. In particular, the most popular three RCPs all involve simultaneously changed functions. 
}
\end{tcolorbox}

\subsubsection{Case Studies for The Three Most Popular RCPs}

\red{We conducted two sets of case studies to understand (1) the semantic meanings of P1--P3 and (2) any co-change indicators within code for those patterns. In each set of case studies, we randomly sampled 20 commits matching each of these RCPs and manually analyzed the code changes in all 60 commits.
}

\paragraph*{\red{\textbf{The Semantic Meanings of P1--P3}}} \red{In the 20 commits sampled for each pattern, we summarized the semantic relevance of entity-level changes as below.}

\red{\textbf{Observations for P1 (*CF$\xrightarrow{f}$CF)}. We found the caller and callee functions changed together in three typical scenarios. First, in about 45\% of the inspected commits, both caller and callee functions experienced \emph{consistent changes} to invoke the same function(s), access the same variable(s), or execute the same statement(s). Second, in about 40\% of the commits, developers applied \emph{adaptive changes} to callers when callees were modified. The adaptive changes involve modifying caller implementations when the signatures of callee functions were updated, or moving code from callees to callers. Third, in 15\% of cases, callers and callees experienced seemingly irrelevant changes.}

\red{\textbf{Observations for P2 (*CF$\xrightarrow{f}$AF).} Such changes were applied for two major reasons. First, in 65\% of the studied commits, the added function implemented some new logic, which was needed by the changed caller function. Second, in the other 35\% of cases, changes were applied for refactoring purposes. Namely, the added function was extracted from one or more existing functions and those functions were simplified to just invoke the added function. }

\red{\textbf{Observations for P3 (*CF$\xrightarrow{v}$AV).} Developers applied such changes in two typical scenarios. First, in 60\% of cases, developers added a new variable for feature addition, which variable was needed by each changed function (i.e., cross-cutting concern~\cite{Ingeno18}). Second, in 40\% of the cases, developers added variables for refactoring purposes. For instance, some developers added a variable to replace a whole function, so all caller functions of the replaced function were consistently updated to instead access the new variable. Additionally, some other developers added a variable to replace some expression(s), constant(s), or variable(s). Consequently, the functions related to the replaced entities were consistently updated for the new variable. }

\paragraph*{\red{\textbf{The Code Indicators for Co-Changes in P1--P3}}}

\red{To identify potential ways of recommending changes based on the mined RCPs, we randomly picked 20 commits matching each pattern among P1--P3;} we ensured that each sampled commit has two or more co-changed functions (e.g., *CF) referencing another edited entity.
 We then inspected the co-changed functions
 in each commit, to decide whether they share any commonality that may indicate their simultaneous changes.
As shown in Table~\ref{tab:common}, the three case studies I--III correspond to the three patterns P1--P3 in sequence. In our manual analysis, we mainly focused on four types of commonality:

\begin{itemize}
\item \textbf{FI}: The co-changed functions commonly invoke one or more \emph{peer functions} of the depended entity $E$ (i.e., CF in P1, AF in P2, and AV in P3). Here, \textbf{peer function} is any function that is defined in the same file as $E$.

\item \textbf{VA}: The co-changed functions commonly access one or more \emph{peer variables} of the depended entity $E$. Here, \textbf{peer variable} is any variable that is defined in the same file as $E$.

\item \textbf{ST}: The co-changed functions commonly share at least 50\% of their token sequences. We calculated the percentage with the longest common subsequence algorithm between two token strings.

\item \textbf{SS}: The co-changed functions commonly share at least 50\% of their statements. We computed the percentage by recognizing identical statements between two given functions \codefont{f1(...)} and \codefont{f2(...)}. Assume that the two functions separately contain $n_1$ and $n_2$ statements, and the number of common statements is $n_3$. Then the percentage is calculated as
\begin{equation}
 \dfrac{n_3 \times 2}{{n_1 + n_2}} \times 100\%
\end{equation}
\end{itemize}

\begin{table}[h]
\caption{Commonality observed between the co-changed functions}
\label{tab:common}
\scriptsize
\centering
\begin{tabular}{R{3cm}| R{1.2cm} R{1.2cm} R{1.2cm} R{1.2cm} |R{2cm}}
\toprule
\multirow{2}{*}{\textbf{Case Study}} & \multicolumn{4}{c|}{\textbf{Commonality}} & {\textbf{No}}\\ \cline{2-5}
&  \textbf{FI} & \textbf{VA} & \textbf{ST} & \textbf{SS} & \textbf{Commonality} \\ \hline
I (for P1: *CF$\xrightarrow{f}$CF) &8 &5 &7 &4 &4 \\ \hline
II (for P2: *CF$\xrightarrow{f}$AF) &12 & 7 & 8 &6 &2 \\ \hline
III (for P3: *CF$\xrightarrow{v}$AV) & 6 & 13 & 6 &5 &3\\
\bottomrule
\end{tabular}
\end{table}

According to Table~\ref{tab:common}, 80\%--90\% of co-changed functions share certain commonality with each other. There are only 2--4 commits in each study where the co-changed functions share nothing in common. Particularly, in the first case study, the FI commonality exists in eight commits, VA exists in five commits, ST occurs in seven commits, and SS occurs in four commits. The summation of these commonality occurrences is larger than 20, because the co-changed functions in some commits share more than one type of commonality.
Additionally, the occurrence rates of the four types of commonality are different between case studies. For instance, FI has 8 occurrences in the first case study; it occurs in 12 commits of the second study and occurs in only 6 commits of the third study. As another example, most commits (i.e., 13) in the third study share the VA commonality, while there are only 5 commits in the first study having such commonality. The observed differences between our case studies imply that when developers apply multi-entity edits matching different RCPs, the commonality shared between co-changed functions also varies.

\begin{tcolorbox}
	\textbf{Finding 4:} \emph{When inspecting the relationship between co-changed functions in three case studies, we found that these functions usually share certain commonality. This indicates great opportunities for developing co-change recommendation approaches.
}
\end{tcolorbox} 
\section{Our Change Recommendation  Approach: \tool}
\label{sec:approach}

\begin{figure}
\includegraphics[width=\linewidth]{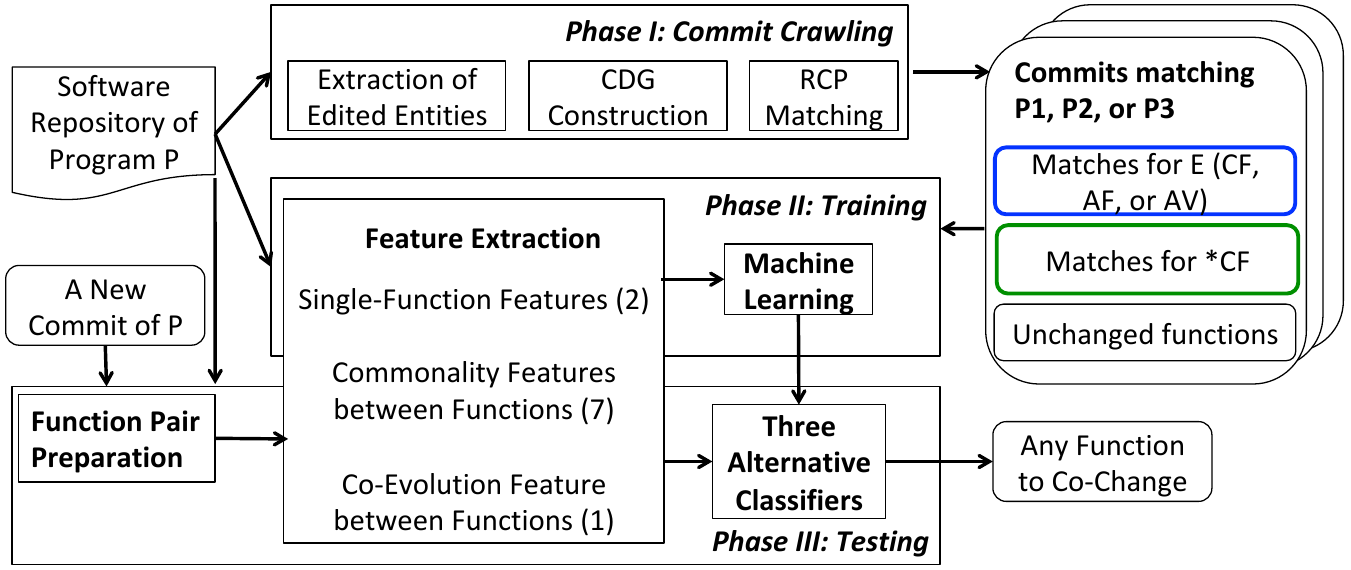}
\caption{\tool consists of three phases: commit crawling, training, and testing}\label{fig:overview}
\end{figure}

In our characterization study (see Section~\ref{sec:study}), we identified three most popular RCPs: *CF$\xrightarrow{f}$CF, *CF$\xrightarrow{f}$AF, and *CF$\xrightarrow{v}$AV. In all these patterns, there is at least one or more changed functions (i.e., *CF) that references another edited entity $E$ (i.e., CF, AF, or AV). In the scenarios when two or more co-changed functions commonly depend on $E$, we also observed certain commonality between those functions.
This section introduces our recommendation system---\tool---which is developed based on the above-mentioned insights. As shown in Figure~\ref{fig:overview}, \tool has three phases.
In the following subsections (Sections~\ref{sec:phase1}-\ref{sec:phase3}), we explain each phase in detail. 

\subsection{Phase I: Commit Crawling}\label{sec:phase1}
Given the repository of a project $P$, Phase I crawls commits to locate any data usable for machine learning.
Specifically, for each commit in the repository, this phase reuses part of our study approach (see Sections~\ref{sec:extract} and~\ref{sec:cdg}) to extract edited entities and to create CDGs. If a commit $c$ has any subgraph matching P1, P2, or P3, this phase recognizes the entity $E_m$ matching $E$ (i.e., an entity matching CF in P1, matching AF in P2, or matching AV in P3) and any co-changed function matching *CF. We denote these co-changed function(s) with $CF\_Set=\{cf_1, cf_2, \ldots\}$, and denote the unchanged function(s) in edited JS files from the same commit with $UF\_Set=\{uf_1, uf_2, \ldots\}$. If $CF\_Set$ has at least two co-changed functions, \tool considers the commit to be usable for model training and passes $E_m$, $CF\_Set$, and $UF\_Set$ to the next phase.


\subsection{Phase II: Training}\label{sec:phase2}
This phase has two major inputs: the software repository of program P, and the extracted data from each relevant commit (i.e., $E_m$, $CF\_Set$, and $UF\_Set$). In this phase,
\tool first creates positive and negative training samples, and then extracts features for each sample. Next, \tool trains a machine learning model by applying 
Adaboost (with Random Forest as the ``weak learner'')~\cite{adaboost}
to the extracted features.
Specifically,
 to create positive samples, \tool enumerates all possible function pairs in $CF\_Set$, because each pair of these functions were co-changed with $E_m$. We represent the positive samples with $Pos=\{(cf_1, cf_2), (cf_2, cf_1), (cf_1, cf_3), \ldots\}$.
 To create negative samples, \tool pairs up each changed function $cf\in CF\_Set$ with an unchanged function $uf\in UF\_Set$, because each of such function pairs were not co-changed. Thus, we denote the negative samples as $Neg=\{(cf_1, uf_1), (uf_1, cf_1), (cf_1, uf_2), \ldots\}$.
 By preparing positive and negative samples in this way, given certain pair of functions, we expect the trained model to predict whether the functions should be co-changed or not.

 \begin{table}
	\centering
	\caption{A list of features extracted for function pair $(f_1, f_2)$}
	\label{tab:features}
	\scriptsize
	\begin{tabular}{R{0.5cm}|L{4.8cm}||R{0.5cm}| L{4.8cm}} \toprule
	\textbf{Id} & \textbf{Feature} & \textbf{Id} & \textbf{Feature} \\ \toprule
	1 & Number of $E_m$-relevant parameter types in $f_2$& 6 & Whether $f_1$ and $f_2$ have the same return type \\ \hline
	2 & Whether $f_2$ has the $E_m$-related type & 7 &Whether $f_1$ and $f_2$ are defined in the same way\\ \hline
	3 & Number of common peer variables & 8 &Token similarity \\ \hline
	4 & Number of common peer functions & 9 & Statement similarity\\ \hline
	5 & Number of common parameter types & 10 & Co-evolution frequency \\ \bottomrule
	\end{tabular}
	\vspace{-0.5em}
\end{table}

\tool extracts 10 features for each sample. As illustrated in Figure~\ref{fig:overview}, two features reflect code characteristics of the second function in the pair, seven features capture the code commonality between functions, and one feature focuses on the co-evolution relationship between functions. Table~\ref{tab:features} presents more details of each feature. Specifically, the $1^{st}$ and $2^{nd}$ features are about the relationship between $f_2$ and $E_m$. Their values are calculated as below: 
\begin{itemize}
\item When $E_m$ is CF or AF, the $1^{st}$ feature records the number of types used in $f_2$ that match any declared parameter type of $E_m$. Intuitively, the more type matches, the more likely that $f_2$ should be co-changed with $E_m$. The $2^{nd}$ feature checks whether the code in $f_2$ uses the return type of $E_m$. 
\item When $E_m$ is AV, the $1^{st}$ feature is set to zero, because there is no parameter type involved in variable declaration. The $2^{nd}$ feature checks whether the code in $f_2$ uses the data type of the newly added variable. 

\end{itemize}

The $3^{rd}$ and $4^{th}$ features were calculated in similar ways. Specifically, depending on which JS file defines $E_m$, \tool locates peer variables (i.e., variables defined within the same file as $E_m$) and peer functions (i.e., functions defined in the same file). Next, \tool identifies the accessed peer variables (or peer functions) by each function in the pair, and intersects the sets of both functions to count the commonly accessed peer variables (or peer functions). 
Additionally, 
 the $7^{th}$ feature checks whether $f_1$ and $f_2$ are defined in the same manner. In our research, we consider the following five ways to define functions: 
\begin{itemize}
\item[(1)] via \codefont{FunctionDeclaration}, e.g., \codefont{function foo(...)\{...\}}, 
\item[(2)] via \codefont{VariableDeclaration}, e.g., \codefont{var foo = function(...)\{...\}}, 
\item[(3)] via \codefont{MethodDefinition}, e.g., \codefont{Class A \{foo(...)\{...\}\}},
\item[(4)] via \codefont{PrototypeFunction} to extend the prototype of an object or a function, e.g., \codefont{x.prototype.foo = function(...)\{...\}}, and
{\item[(5)] via certain \codefont{exports}-related statements, e.g., \codefont{exports.foo = function(...)\{...\}} and \codefont{module.exports = \{foo: function(...)\{...\}\}}.}
\end{itemize}
If $f_1$ and $f_2$ are defined in the same way, the $7^{th}$ feature is set to \codefont{true}. 
Finally, the $10^{th}$ feature assesses in the commit history, how many times the pair of functions were changed together before the current commit. Inspired by prior work~\cite{Zimmermann:2004}, we believe that the more often two functions were co-changed in history, the more likely that they are co-changed in the current or future commits.

Depending on the type of $E_m$, \tool takes in extracted features to actually train three independent classifiers, with each classifier corresponding to one pattern among P1--P3. For instance, one classifier corresponds to P1: *CF$\xrightarrow{f}$CF. Namely, when $E_m$ is CF and one of its caller functions $cf$ is also changed, this classifier predicts whether there is any unchanged function $uf$ that invokes $E_m$ and should be also changed. The other two classifiers separately predict functions for co-change based on P2 and P3. We consider these three binary-class classifiers as an integrated machine learning model, because all of them can take in features from one program commit and related software version history, in order to recommend co-changed functions when possible.

\subsection{Phase III: Testing}\label{sec:phase3}

This phase takes in two inputs---a new program commit $c_n$ and the related software version history, and recommends any unchanged function that should have been changed by that commit. Specifically, given $c_n$, \tool reuses the steps of Phase I (see Section~\ref{sec:phase1}) to locate $E_m$, $CF\_Set$, and $UF\_Set$. \tool then pairs up every changed function $cf\in CF\_Set$ with every unchanged one $uf\in UF\_Set$, obtaining a set of candidate function pairs $Candi=\{(cf_1, uf_1), (uf_1, cf_1), (cf_1, uf_2), \ldots\}$. Next, \tool extracts features for each candidate $p$ and sends the features to a pre-trained classifier depending on $E_m$'s type. If the classifier predicts the function pair to have co-change relationship, \tool recommends developers to also modify the unchanged function in $p$.

\section{Evaluation}
\label{sec:eval}

\red{In this section, we first introduce our experiment setting (Section~\ref{sec:data}) and the metrics used to evaluate \tool's effectiveness (Section~\ref{sec:metrics}). Then we explain our investigation with different ML algorithms and present \tool's sensitivity to the adopted ML algorithms (Section~\ref{sec:compare_models}), through which we finalize the default ML algorithm applied in \tool.  Next we expound on the effectiveness comparison between \tool and two existing tools: ROSE~\cite{Zimmermann:2004} and
Transitive Associate Rules (TAR)~\cite{Islam:2018} (Section~\ref{sec:compare}). 
	Finally, we present the comparison between \tool and a variant approach that trains one unified classifier instead of three distinct classifiers to recommend changes (Section~\ref{sec:variant}). }

\subsection{Experiment Setting}
\label{sec:data}

\begin{table}
\centering
\scriptsize
\caption{Numbers of commits that are potentially usable for model training and testing}
\label{tab:1a1c}
	\begin{tabular}{L{2.5cm}|R{2.5cm} R{2.5cm} R{2.5cm} }
		\toprule
		\textbf{Project} & \textbf{\# of Commits Matching P1} & \textbf{\# of Commits Matching P2} & \textbf{\# of Commits Matching P3} 
		\\
		\toprule
Node.js & 92 & 77 & 65 \\ \hline
Meteor & 67 & 59 & 39  \\ \hline
Ghost & 21 & 24 & 28 \\ \hline
Habitica  & 11 & 8 & 5 \\ \hline
PDF.js & 14 & 12 & 14 \\ \hline
React & 18 & 12 & 5 \\ \hline
Serverless & 26 & 12 & 8 \\ \hline
Webpack & 22 & 24 & 8 \\ \hline
Storybook & 2 & 1 &4 \\ \hline
Electron & 7 & 3 & 6 \\
\bottomrule
\textbf{Sum} & \textbf{280} & \textbf{232} & \textbf{182} \\
\bottomrule
\end{tabular}
\end{table}

We mined repositories of the 10 open-source projects introduced in Section~\ref{sec:study}, and found three distinct sets of commits in each project that are potentially usable for model training and testing. As shown in Table~\ref{tab:1a1c}, in total, we found 280 commits matching P1, 232 commits matching P2, and 182 commits matching P3. Each of these pattern matches has at least two co-changed functions (*CF) depending on $E_m$. In our evaluation, for each data set of each project, we could use part of the data to train a classifier and use the remaining data to test the trained classifier. Because Storybook and Electron have few commits, we excluded them from our evaluation and simply used the identified commits of the other eight projects to train and test classifiers. 

\begin{figure}
\centering
\includegraphics[width=0.85\linewidth]{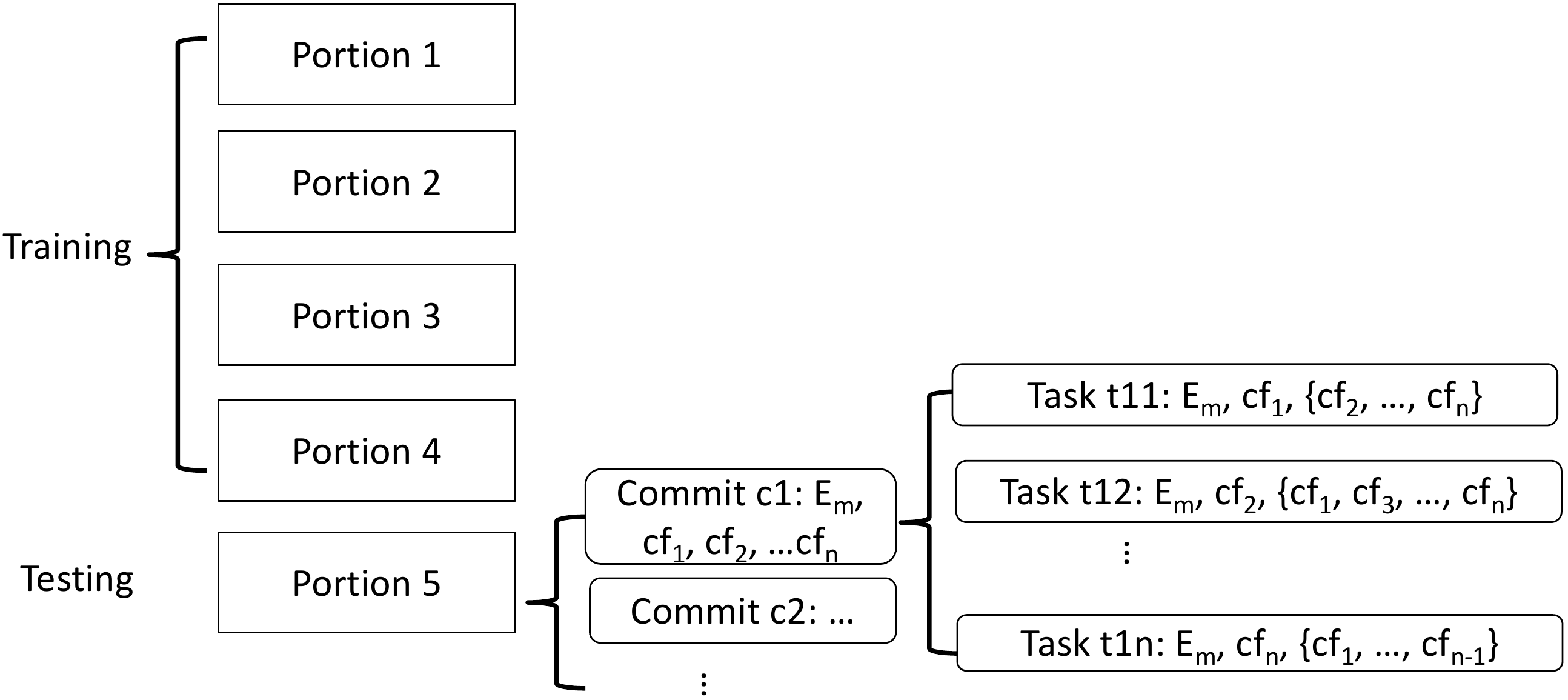}
\caption{Typical data processing for each fold of the five-fold cross validation}
\label{fig:cross-validation}
\end{figure}
\red{We decided to use $k$-fold cross validation to evaluate \tool's effectiveness. Namely,  for every data set of each project, we split the mined commits into $k$ portions roughly evenly; each fold uses $(k-1)$ data portions for training and the remaining portion for testing. Among the eight projects, because each project has at least five commits matching each pattern, we set $k=5$ to diversify our evaluation scenarios as much as possible. For instance, Habitica has five commits matching P3. When evaluating \tool's capability of predicting co-changes for Hibitica based on P3, in each of the five folds, we used four commits for training and one commit for testing.}
\red{Figure~\ref{fig:cross-validation} illustrates our five-fold cross validation procedure. In the procedure, we ensured that each of the five folds adopted a distinct data portion to construct prediction tasks for testing purposes. For any testing commit that has $n$ co-changed functions (*CF) depending on $E_m$, i.e., $CF\_Set=\{cf_1, cf_2, \ldots, cf_n\}$, we created $n$ prediction tasks in the following way. In each prediction task, we included one known changed function $cf_i$ $(i \in [1, n])$ together with $E_m$ and kept all the other $(n-1)$ functions unchanged. We regarded the $(n-1)$ functions as ground truth ($GT$) to assess how accurately \tool can recommend co-changes given $E_m$ and $cf_i$. } 

\red{For instance, one prediction task we created in React includes the followings: $E_m$ = \codefont{src/isomorphic/classic/types.ReactPropTypes.createChainableTypeChecker(...)}, $cf$ = \codefont{src/isomorphic/classic/types.ReactPropTypes.createObjectOfTypeChecker(...)}, and  $GT = $\{\codefont{src/isomorphic/ classic/types.ReactPropTypes.createShapeType\-Checker(...)}\}. When \tool blindly pairing $cf$ with any unchanged function, it may extract feature values as below: feature1 = 1, feature2 = \codefont{true}, feature3 = 0, feature4 = 2, feature5 = 0, feature6 = \codefont{true}, feature7 = \codefont{true}, feature8 = 76\%, feature9 = 45\%, feature10 = 1\}. Table~\ref{tab:instances} shows the total numbers of prediction tasks we created for all projects and all patterns among the five-fold cross validation.}


\begin{table}
	\centering
	\scriptsize
	\caption{Total numbers of prediction tasks involved in the five-fold cross validation}
	\label{tab:instances}
	\begin{tabular}{L{2.5cm}|R{2.5cm} R{2.5cm} R{2.5cm} }
		\toprule
		\textbf{Project} & \textbf{\# of Tasks Matching P1} & \textbf{\# of Tasks Matching P2} & \textbf{\# of Tasks Matching P3} 
		\\
		\toprule
		Node.js & 398 & 309 & 223 \\ \hline
		Meteor & 401 & 229 & 107  \\ \hline
		Ghost & 76 & 77 & 99 \\ \hline
		Habitica  & 30 & 23 & 18 \\ \hline
		PDF.js & 41 & 31 & 35 \\ \hline
		React & 72 & 37 & 17 \\ \hline
		Serverless & 81 & 38 & 23 \\ \hline
		Webpack & 138 & 90 & 22 \\ 
		\bottomrule
		\textbf{Sum} & \textbf{1,237} & \textbf{834} & \textbf{544} \\
		\bottomrule
	\end{tabular}
\end{table}

\subsection{Metrics}
\label{sec:metrics}
We defined and used four metrics to measure a tool's capability of recommending co-changed functions: coverage, precision, recall, and F-score. We also defined the weighted average to measure a tool's overall effectiveness among all subject projects for each of the metrics mentioned above.

\textbf{Coverage (Cov)} is the percentage of tasks for which a tool can provide suggestion.
Given a task, a tool may or may not recommend any change to complement the already-applied edit, so
this metric assesses the tool applicability. Intuitively, the more tasks for which a tool can recommend one or more changes, the more applicable this tool is.

\begin{equation}
\small
Cov = \frac{\text{\# of tasks with a tool's suggestion}}{\text{Total \# of tasks}} \times 100\%
\vspace{-0.2em}
\end{equation}
Coverage varies within [0\%, 100\%]. If a tool always recommends some change(s) given a task, its coverage is 100\%.
All our later evaluations for precision, recall, and F-score are limited to the tasks covered by a tool. For instance, suppose that given 100 tasks, a tool can recommend changes for 10 tasks. Then the tool's coverage is $10/100=10\%$, and the evaluations for other metrics are based on the 10 instead of 100 tasks.

\textbf{Precision (Pre)} measures among all recommendations by a tool, how many of them are correct:
\begin{equation}
\small
Pre = \frac{\text{\# of correct recommendations}}{\text{Total \# of recommendations by a tool}} \times 100\%
\end{equation}
\noindent
This metric evaluates how precisely a tool recommends changes.
If all suggestions by a tool are contained by the ground truth, the precision is 100\%.

\textbf{Recall (Rec)} measures among all the expected recommendations, how many of them are actually reported by a tool:
\begin{equation}
\small
Rec = \frac{\text{\# of correct recommendations by a tool}}{\text{Total \# of expected recommendations}} \times 100\%
\end{equation}
This metric assesses how effectively a tool retrieves the expected co-changed functions. Intuitively, if all expected recommendations are reported by a tool, the recall is 100\%.

\textbf{F-score (F1)} measures the accuracy of a tool's recommendation:
\begin{equation}
\small
F1 = \frac{2 \times Pre \times Rec}{Pre + Rec}\times 100\%
\end{equation}
F-score is the harmonic mean of precision and recall. Its value varies within [0\%, 100\%].
The higher F-scores are desirable, as they demonstrate better trade-offs between the precision and recall rates.

\textbf{Weighted Average (WA)} measures a tool's \textbf{overall effectiveness} among all experimented data in terms of coverage, precision, recall, and F-score:
\begin{equation}
\small
	\Gamma_{overall} = \frac{\sum_{i=1}^{8} \Gamma_i*n_i}{\sum_{i=1}^{8} n_i}.
\label{eqoverallthree}
\end{equation}
In the formula, $i$ varies from 1 to 8, representing the 8 projects used in our evaluation (Storybook and Electron were excluded). Here, $i=1$ corresponds to Node.js and $i=8$ corresponds to Webpack; $n_i$ represents the number of tasks built from the $i^{th}$ project.
$\Gamma_i$ represents any measurement value of the $i^{th}$ project for coverage, precision, recall, or F-score.
By combining such measurement values of eight projects in a weighted way, we were able to assess a tool's overall effectiveness $\Gamma_{overall}$.

{\subsection{Sensitivity to The Adopted ML Algorithm}
\label{sec:compare_models}
\red{We designed \tool to use Adaboost, with Random Forests as the weak learners to train classifiers. To make this design decision, we tentatively integrated \tool with five alternative algorithms: J48~\cite{j48}, Random Forest~\cite{RandomForest}, Na\"ive Bayes~\cite{nb}, Adaboost (default), and Adaboost (Random Forest).}

\begin{figure}
\begin{subfigure}{\linewidth}
\includegraphics[width=\linewidth]{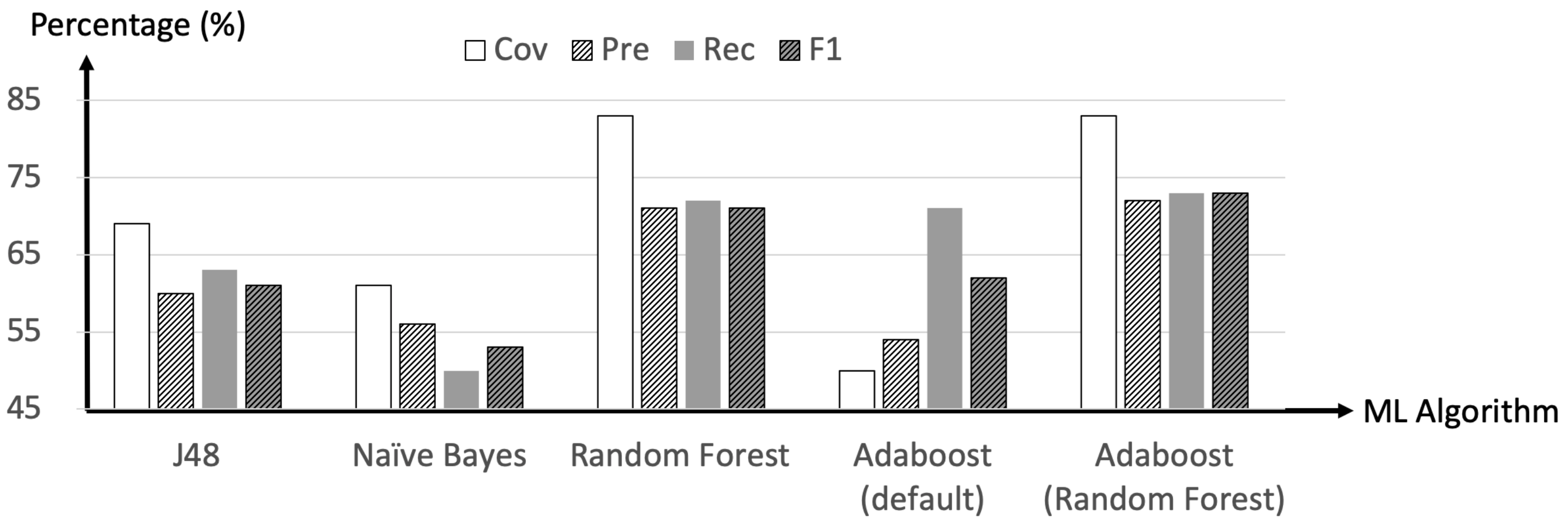}
\vspace{-2.5em}
\caption{The *CF$\xrightarrow{f}$CF data}
\end{subfigure}
\begin{subfigure}{\linewidth}
\includegraphics[width=\linewidth]{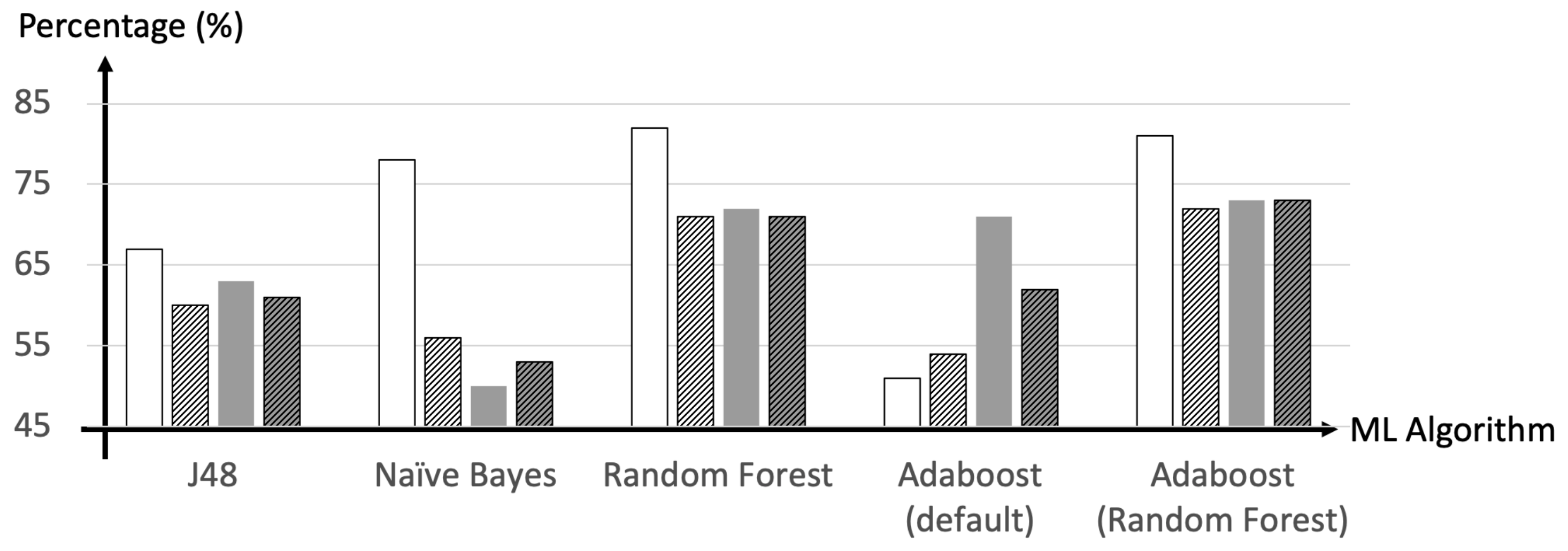}
\vspace{-2.5em}
\caption{The *CF$\xrightarrow{f}$AF data}
\end{subfigure}
\begin{subfigure}{\linewidth}
\includegraphics[width=\linewidth]{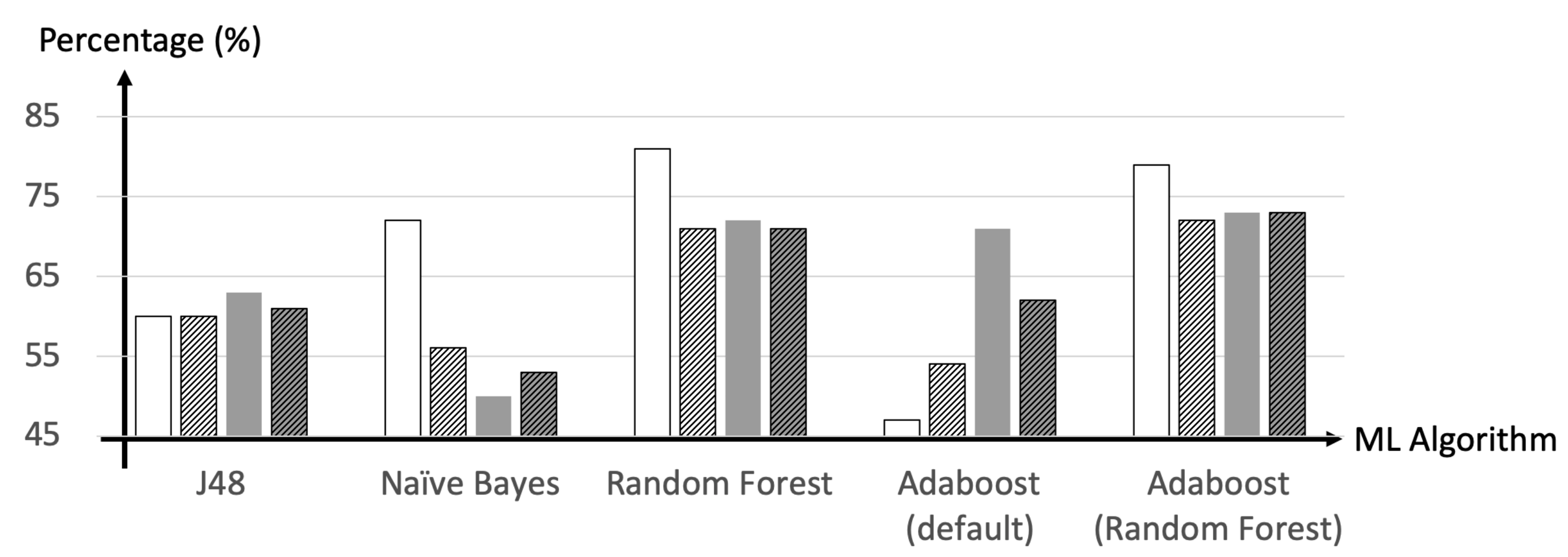}
\vspace{-2.5em}
\caption{The *CF$\xrightarrow{v}$AV data}
\end{subfigure}
\caption{Comparison between different ML algorithms on different data sets}
\label{fig:ml-comparison}
\end{figure}

\begin{itemize}
\item \textbf{J48} builds a decision tree as a predictive model to go from observations about an item (represented in the branches) to conclusions about the item's target value (represented in the leaves).
\item \textbf{Na\"ive Bayes} calculates the probabilities of hypotheses by applying Bayes' theorem with strong (na\"ive) independence assumptions between features.
\item \textbf{Random Forest} is an ensemble learning method that trains a model to make predictions based on a number of different models. Random Forest trains a set of individual models in a parallel way. Each model is trained with a random subset of the data. Given a candidate in the testing set, individual models make their separate predictions and Random Forest uses the one with the majority vote as its final prediction.
\item \textbf{Adaboost} is also an ensemble learning method. However, different from Random Forest, Adaboost trains a bunch of individual models (i.e., weak learners) in a sequential way. Each individual model learns from mistakes made by the previous model. We tried two variants of Adaboost: (1) Adaboost (default) with decision trees as the weak learners, and (2) Adaboost (Random Forest) with Random Forests as the weak learners.
\end{itemize}

Figure~\ref{fig:ml-comparison} illustrates the effectiveness comparison when \tool adopts different ML algorithms. The three subfigures (Figure~\ref{fig:ml-comparison} (a)--(c)) separately present the comparison results on the data sets of *CF$\xrightarrow{f}$CF, *CF$\xrightarrow{f}$AF, and *CF$\xrightarrow{v}$AV. We observed similar phenomena in all subfigures.  \red{By comparing the first four basic ML algorithms (J48, Na\"ive Bayes, Random Forest, and Adaboost (default)), we noticed that Random Forest achieved the best results in all metrics. Among all datasets, Na\"ive Bayes obtained the lowest recall and accuracy rates. Although Adaboost obtained the second highest F-score, its coverage is the lowest probably because it uses decision trees as the default weak learners. Based on our evaluation with the first four basic algorithms, we were curious how well Adaboost performs if it integrates Random Forests as weak learners. Thus, we also experimented with a fifth algorithm: Adaboost (Random Forest).}

\red{As shown in Figure~\ref{fig:ml-comparison}, Adaboost (Random Forest) and Random Forest achieved very similar effectiveness, and both of them considerably outperformed the other algorithms. But compared with Random Forest, Adaboost (Random Forest) obtained better precision, better recall, better accuracy, and equal or slightly lower coverage. Thus, we chose Adaboost (Random Forest) as the default ML algorithm used in \tool.}
Our results imply that although ensemble learning methods generally outperform other ML algorithms, their effectiveness also depends on (1) what weak learners are used and (2) how we organize weak learners. Between Adaboost (Random Forest) and Adaboost (default), the only difference exists in the used weak learner (Random Forest vs.~Decision Tree). Our evaluation shows that Random Forest helps improve Adaboost's performance when it is used as the weak learner. Additionally, between Random Forest and Adaboost (default), the only difference is how they combine decision trees as weak learners. Our evaluation shows that Random Forest outperforms Adaboost by
training weak learners in a parallel instead of sequential way.

\begin{tcolorbox}
	\textbf{Finding 5:}
	\emph{\tool is sensitive to the adopted ML algorithm. \tool obtained the lowest prediction accuracy when Na\"ive Bayes was used, but acquired the highest accuracy when Adaboost (Random Forest) was used.
	}
\end{tcolorbox}

}

\subsection{Effectiveness Comparison with ROSE \red{and TAR}}
\label{sec:compare}

In our evaluation, we compared \tool with a popularly used tool ROSE~\cite{Zimmermann:2004} \red{and a more recent tool Transitive Associate Rules (TAR)~\cite{Islam:2018}. Both of these tools recommend changes by mining co-change patterns from version history.} 

\red{Specifically,} ROSE mines the association rules between co-changed entities from software version history. An exemplar mined rule is shown below:

\begin{equation}
\small
\begin{array}{l}
\{(\_Qdmodule.c, func, GrafObj\_getattr())\}\Rightarrow \\
\left\{
\begin{array}{l}
(qdsupport.py, func, outputGetattrHook()).\\
\end{array}
\right\}\\
\end{array}
\end{equation}
This rule means that whenever the function \codefont{GrafObj\_getattr()} in a file {\sf \_Qdmodule.c} is changed, the function \codefont{outputGetattrHook()} in another file {\sf qdsupport.py} should also be changed. Based on such rules, given a program commit, ROSE tentatively matches all edited entities with the antecedents of all mined rules and recommends co-changes if any tentative match succeeds. 
\red{Similar to ROSE, TAR also mines association rules from version history. However, in addition to the mined rules (e.g., $E1 \Rightarrow E2$ and $E2 \Rightarrow E3$), TAR also leverages \textbf{transitive inference} to derive more rules (e.g., $E1 \Rightarrow E3$); it computes the confidence value of each derived rule based on the confidence values of original rules (e.g., $conf(E1\Rightarrow E3) = conf(E1 \Rightarrow E2) \times conf(E2 \Rightarrow E3)$).} 

In our comparative experiment, we applied ROSE and TAR to the constructed prediction tasks and version history of each subject project. We configured ROSE with $support=1$ and $confidence=0.1$, because
the ROSE paper~\cite{Zimmermann:2004} mentioned this setting multiple times \red{and it achieved the best results by balancing recall and precision. For consistency, we also configured TAR with $support=1$ and $confidence=0.1$.}

As shown in Table~\ref{tab:1cf1c}, \tool outperformed ROSE \red{and TAR} in terms of all measurements. Take Webpack as an example. Among the 138 {*CF$\xrightarrow{f}$CF} prediction tasks in this project,  
\tool provided change recommendations for 89\% of tasks; with these recommendations, \tool achieved 71\% precision, 81\% recall, and 75\% accuracy. On the other hand, ROSE \red{and TAR} recommended changes for only 50\% of tasks; based on its recommendations, ROSE acquired only 7\% precision, 29\% recall, and 12\% accuracy\red{, while TAR obtained 5\% precision, 34\% recall, and 9\% accuracy.}
Among the eight subject projects, the weighted average measurements of \tool include 83\% coverage, 72\% precision, 73\% recall, and 73\% accuracy. Meanwhile, the weighted average measurements of ROSE include 53\% coverage, 21\% precision, 52\% recall, and 29\% accuracy. \red{TAR achieved 59\% average recall, but its average precision and accuracy are the lowest among the three tools}.  Such measurement contrasts indicate that \tool usually recommended more changes than ROSE \red{or TAR}, and \tool's recommendations were more accurate. 

\begin{table}
	\centering
	\scriptsize
	\caption{Evaluation results of \tool, ROSE, and TAR
		for {*CF$\xrightarrow{f}$CF} tasks (\%)}
	\label{tab:1cf1c}
	\begin{tabular}{C{1.2cm}| R{0.4cm} R{0.4cm} R{0.4cm} R{0.35cm}
	 | R{0.4cm} R{0.4cm} R{0.4cm} R{0.35cm} | R{0.4cm} R{0.4cm} R{0.4cm} R{0.35cm}
	}
		\toprule
		\multirow{2}{*}{\textbf{Project}} & \multicolumn{4}{c|}{\textbf{\tool}} & \multicolumn{4}{c|}{\textbf{ROSE}} & \multicolumn{4}{c}{\textbf{TAR}}\\
		\cline{2-13}
		& \textbf{Cov} & \textbf{Pre} & \textbf{Rec} & \textbf{F1} & \textbf{Cov} & \textbf{Pre} & \textbf{Rec} & \textbf{F1} & \textbf{Cov} & \textbf{Pre} & \textbf{Rec} & \textbf{F1} \\
		\toprule
		Node.js & 77 & 68 & 69 & 69 & 61 & 24 & 56& 34 & 65 & 15 & 62 & 24\\
		Meteor & 88 & 72 & 70 & 71 & 46 & 16 & 43 & 24 & 52 & 15 & 47 & 23\\
		Ghost & 73 & 67 & 74 & 71  & 50 & 20 & 53 & 29 & 50 & 14 & 57 & 22\\
		Habitica & 80 & 80 & 78 & 79 & 40 & 7 & 37 & 12 & 35 & 5 & 42 & 9\\
		PDF.js & 71 & 77 & 81 & 79 & 29 & 27 & 41 & 33 & 33 & 8 & 45 & 14\\
		React & 91 & 86 & 76 & 81 & 32 & 59 & 70 & 64 & 32 & 57 & 74 & 64\\
		Serverless & 84 & 77 & 79 & 78  & 64& 20 & 75 & 32 & 68 & 16 & 80 & 27\\
		Webpack & 89 & 71 & 81 & 75 & 50 & 7 & 29 & 12 & 50 & 5 & 34 & 9\\
		\bottomrule
		\textbf{WA} & \textbf{83} & \textbf{72} & \textbf{73} & \textbf{73} & \textbf{53} & \textbf{21} & \textbf{52} & \textbf{29} & \textbf{57} & \textbf{15} & \textbf{59} & \textbf{24}\\
		\bottomrule
	\end{tabular}
\end{table}

\begin{table}
	\centering
	\scriptsize
	\vspace{1em}
	\caption{Result comparison among \tool, ROSE, and TAR for {*CF$\xrightarrow{f}$AF} tasks (\%)}
	\label{tab:1af1c}
	\begin{tabular}{C{1.2cm}|R{0.4cm} R{0.4cm} R{0.4cm} R{0.35cm} | R{0.4cm} R{0.4cm} R{0.4cm} R{0.35cm} | R{0.4cm} R{0.4cm} R{0.4cm} R{0.35cm}}
		\toprule
		\multirow{2}{*}{\textbf{Project}} & \multicolumn{4}{c|}{\textbf{\tool}} & \multicolumn{4}{c|}{\textbf{ROSE}} & \multicolumn{4}{c}{\textbf{TAR}}\\
		\cline{2-13}
		& \textbf{Cov} & \textbf{Pre} & \textbf{Rec} & \textbf{F1} & \textbf{Cov} & \textbf{Pre} & \textbf{Rec} & \textbf{F1} & \textbf{Cov} & \textbf{Pre} & \textbf{Rec} & \textbf{F1} \\
		\toprule
		Node.js & 79 & 69 & 74 & 72 & 59 & 20 & 52 & 29 & 61 & 14 & 61 & 23\\
		Meteor & 86 & 77 & 82 & 80 & 40 & 22 & 44 & 29 & 46 & 21 & 50 & 29\\
		Ghost & 85 & 86 & 85 & 85  & 46 & 18 & 46 & 26 & 50 & 14 & 49 & 22\\
		Habitica & 87 & 77 & 85 & 81 & 56 & 4 & 23 & 7 & 58 & 2 & 39 & 4\\
		PDF.js & 65 & 87 & 88 & 87 & 22 & 9 & 28 & 14 & 23 & 11 & 58 & 19\\
		React & 71 & 84 & 82 & 83 & 16 & 66 & 7 & 13 & 17 & 67 & 8 & 14\\
		Serverless & 84 & 71 & 85 & 77  & 73& 19 & 59 & 29 & 74 & 15 & 60 & 24\\
		Webpack & 75 & 79 & 85 & 82 & 53 & 16 & 46 & 24 & 56 & 13 & 49 & 21\\
		\bottomrule
		\textbf{WA} & \textbf{81} & \textbf{76} & \textbf{80} & \textbf{78} & \textbf{54} & \textbf{21} & \textbf{48} & \textbf{28} & \textbf{56} & \textbf{16} & \textbf{55} & \textbf{24}\\
		\bottomrule
	\end{tabular}
\end{table}

In addition to {*CF$\xrightarrow{f}$CF} tasks, we also compared \tool with ROSE \red{and TAR} for {*CF$\xrightarrow{f}$AF} and {*CF$\xrightarrow{v}$AV} tasks, as shown in Tables ~\ref{tab:1af1c} and ~\ref{tab:1av1c}. 
Similar to what we observed in Table~\ref{tab:1cf1c}, \tool outperformed ROSE \red{and TAR} in terms of all metrics for both types of tasks. As shown in Table~\ref{tab:1af1c}, given *CF$\xrightarrow{f}$AF tasks, on average, \tool achieved 81\% coverage, 76\% precision, 80\% recall, and 78\% accuracy. ROSE acquired 54\% coverage, 21\% precision, 48\% recall, and 28\% accuracy\red{. TAR obtained 56\% coverage, 16\% precision, 55\% recall, and 24\% accuracy}. In Table~\ref{tab:1av1c}, 
for Serverless, \tool achieved 70\% coverage, 80\% precision, 85\% recall, and 82\% accuracy. Meanwhile, ROSE only provided recommendations for 34\% of the tasks, and none of these recommendations is correct. TAR only provided recommendations for 38\% of tasks; with the recommendations, TAR achieved 1\% precision, 13\% recall, and 2\% accuracy.

\begin{table}
	\centering
	\scriptsize
	\vspace{1em}
	\caption{Result comparison among \tool, ROSE, and TAR for {*CF$\xrightarrow{v}$AV} tasks (\%)}
	\label{tab:1av1c}
	\begin{tabular}{C{1.2cm}|R{0.4cm} R{0.4cm} R{0.4cm} R{0.35cm} | R{0.4cm} R{0.4cm} R{0.4cm} R{0.35cm} | R{0.4cm} R{0.4cm} R{0.4cm} R{0.35cm}}
		\toprule
		\multirow{2}{*}{\textbf{Project}} & \multicolumn{4}{c|}{\textbf{\tool}} & \multicolumn{4}{c|}{\textbf{ROSE}} & \multicolumn{4}{c}{\textbf{TAR}}\\
		\cline{2-13}
		& \textbf{Cov} & \textbf{Pre} & \textbf{Rec} & \textbf{F1} & \textbf{Cov} & \textbf{Pre} & \textbf{Rec} & \textbf{F1}  & \textbf{Cov} & \textbf{Pre} & \textbf{Rec} & \textbf{F1} \\
		\toprule
		Node.js & 79 & 72 & 77 & 74 & 55 & 20 & 65 & 31 & 56 & 16 & 74 & 26\\
		Meteor & 72 & 77 & 84 & 81 & 26 & 2 & 14 & 4 & 27 & 2 & 31 & 3\\
		Ghost & 84 & 75 & 81 & 78  & 46 & 18 & 46 & 26 & 38 & 8 & 70 & 14\\
		Habitica & 89 & 82 & 85 & 83 & 27 & 20 & 45 & 28 & 28 & 17 & 54 & 26\\
		PDF.js & 78 & 87 & 84 & 85 & 20 & 4 & 28 & 8 & 20 & 5 & 29 & 8\\
		React & 89 & 73 & 78 & 76 & 36 & 8 & 33 & 13 & 12 & 98 & 34 & 50\\
		Serverless & 70 & 80 & 85 & 82  & 34 & 0 & 0 & - & 38 & 1 & 13 & 2\\
		Webpack & 87 & 86 & 83 & 85 & 36 & 8 & 33 & 13 & 40 & 3 & 34 & 5 \\
		\bottomrule
		\textbf{WA} & \textbf{79} & \textbf{76} & \textbf{81} & \textbf{78} & \textbf{45} & \textbf{17} & \textbf{54} & \textbf{25} & \textbf{47} & \textbf{12} & \textbf{62} & \textbf{19}\\
		\bottomrule
	\end{tabular}
\end{table}

Comparing the results shown in Tables~\ref{tab:1cf1c}--\ref{tab:1av1c}, we found the effectiveness of \tool, ROSE, \red{and TAR} to be stable
across different types of prediction tasks. 
Specifically among the three kinds of tasks, on average, \tool achieved 79\%--83\% coverage, 72\%--76\% precision, 73\%--81\% recall, and 73\%--78\% accuracy. On the other hand, ROSE achieved 45\%--54\% coverage, 17\%--21\% precision, 48\%--54\% recall, and 25\%--29\% accuracy; \red{TAR achieved 47\%--56\% coverage, 12\%--16\% precision, 55\%--62\% recall, and 19\%--24\% accuracy.} The consistent comparison results imply that \tool usually recommended co-changed functions for more tasks, and \tool's recommendations usually had higher quality.  

Two major reasons can explain why \tool worked 
\red{best}. First, ROSE \red{and TAR} purely \red{use} the co-changed entities in version history to recommend changes. 
When the history data is incomplete or some entities were never co-changed before, \red{both tools} may lack evidence to predict co-changes and thus obtain lower coverage and recall rates. 
\red{Additionally, TAR derives more rules than ROSE via transitive inference. Namely, if $E1 \Rightarrow E2$ and $E2 \Rightarrow E3$, then $E1 \Rightarrow E3$. However, it is possible that $E1$ and $E3$ were never co-changed before, neither are they related to each other anyhow. Consequently, the derived rules may contribute to TAR's lower precision.} Meanwhile, \tool extracts nine features from a given commit and one feature from the version history; even though history data provides insufficient indication on the potential co-change relationship between entities, the other features can serve as supplements. 

Second, ROSE \red{and TAR observe} no syntactic or semantic relationship between co-changed entities; thus, \red{they} can infer incorrect rules from co-changed but unrelated entities and achieve lower precision. In comparison, \tool observes the syntactic relationship between co-changed entities by tracing the referencer-referencee relations; it also observes the semantic relationship by extracting features to reflect the commonality (1) between co-changed functions (*CF), and (2) between any changed function $cf$ and the changed entity $E$ on which $cf$ depends ($E$ is CF in P1, AF in P2, and AV in P3). 

Although \tool outperformed ROSE \red{and TAR} in our experiments, we consider \tool as a complementary tool to \red{existing tools}. This is because \tool bases its change recommendations on the three most popular RCPs we found. If some changes do not match any of the RCPs, \tool does not recommend any change but ROSE may suggest some edits. 

\begin{tcolorbox}
	\textbf{Finding 6:}
	\emph{\tool outperformed ROSE \red{and TAR} when predicting co-changed functions based on the three recurring change patterns (P1--P3). \tool serves as a good complementary tool to \red{both tools}.
	}
\end{tcolorbox}

\subsection{Comparison with A Variant Approach}
\label{sec:variant}
Readers may be tempted to train a unified classifier instead of three separate classifiers, because the three classifiers all take in the same format of inputs and output the same types of predictions (i.e., whether to co-change or not). However, as shown in Table~\ref{tab:common}, the commonality characteristics between co-changed functions vary with RCPs. For instance, the co-changed functions in P2 usually commonly invoke peer functions (i.e., FI), the co-changed functions in P3 often commonly read/write peer variables (i.e., VA), and the co-changed functions in P1 have weaker commonality signals for both FI and ST (i.e., common token subsequences). If we mix the co-changed functions matching different patterns to train a single classifier, it is quite likely that the extracted features between co-changed functions become less informative, and the trained classifier has poorer prediction power. 

\begin{table}
	\centering
	\scriptsize
	\vspace{1em}
	\caption{The effectiveness of \toolu when it trains and tests a unified classifier (\%)}
	\label{tab:3patterns}
	\begin{tabular}{C{2.4cm}|R{1.2cm} R{1.2cm} R{1.2cm} R{1.2cm}}
		\toprule
		\multirow{1}{*}{\textbf{Project}} 
		& \textbf{Cov} & \textbf{Pre} & \textbf{Rec} & \textbf{F1}  \\
		\toprule
		Node.js & 72 & 50 & 57 & 53 \\
		Meteor & 77 & 59 & 58 & 59 \\
		Ghost & 53 & 61 & 70 & 65 \\
		Habitica & 55 & 53 & 68 & 60 \\
		PDF.js & 29 & 60 & 73 & 66\\
		React & 76 & 75 & 73 & 74\\
		Serverless & 54 & 47 & 61 & 53 \\
		Webpack & 66 & 54 & 63 & 58\\
		\bottomrule
		\textbf{WA} & \textbf{70} & \textbf{56} & \textbf{61} & \textbf{59}\\
		\bottomrule
	\end{tabular}
\end{table}

To validate our approach design, we also built a variant approach of \tool---\toolu---that trains a unified classifier with the program commits matching either of the three RCPs (P1--P3) and predicts co-change functions with the single classifier. To evaluate \toolu, we clustered the data portions matching distinct RCPs for each project, and conducted five-fold cross validation. As shown in Table~\ref{tab:3patterns}, on average, \toolu recommended changes with 70\% coverage, 56\% precision, 61\% recall, and 59\% accuracy. These measured values are much lower than the weighted averages of \tool reported in Tables~\ref{tab:1cf1c}--\ref{tab:1av1c}. The empirical comparison corroborates our hypothesis that when data matching distinct RCPs are mixed to train a unified classifier, the classifier works less effectively.   

\begin{tcolorbox}
	\textbf{Finding 7:}
	\emph{\toolu worked less effectively than \tool by training a unified classifier with data matching distinct RCPs. This experiment validates our approach design of training three separate classifiers. 
	}
\end{tcolorbox}

\section{Threats to Validity}
\label{sec:threats}

\red{\emph{Threats to External Validity:}}
All our observations and experiment results are limited to the software repositories we used. These observations and results may not generalize well to other JS projects, especially to the projects that use the Asynchronous Module Definition (AMD) APIs to define code modules and their dependencies. 
In the future, we would like to include more diverse projects into our data sets so that our findings are more representative.  


Given a project $P$, \tool adopts commits in $P$'s software version history to train classifiers that can recommend co-changes for new program commits. When the version history has few commits to train classifiers, the applicability of \tool is limited. \tool shares such limitation with existing tools that provide project-specific change suggestions based on software version history~\cite{Zimmermann:2004,Rolfsnes:2016,Islam:2018}. 
To further lower \tool's requirement to available commits in software version history, we plan to investigate more ways to extract features from commits and better capture the characteristics of co-changed functions.   

\red{\emph{Threats to Internal Validity: }In our experiments, we sampled a subset of commits in each project based on the keywords ``bug'', ``fix'', ``error'', ``adjust'', and ``failure'' in commit messages. Our insight is that developers may apply tangled changes (i.e., unrelated or loosely related code changes) in a single commit~\cite{Herzig13}; such commits can introduce data noise and make our research investigation biased. Based on our experience, the commits with above-mentioned keywords are likely to fix bugs, and thus each of such commits may be applied to achieve one maintenance goal and contain no tangled changes. However, the keywords we used may not always accurately locate bug fixes, neither do they guarantee that developers apply no tangled changes in individual sampled commits. In the future, we plan to sample commits in other ways and analyze how our observations vary with the sampling techniques. 
}


\red{\emph{Threats to Construct Validity:}}
When creating recommendation tasks for classifier evaluation, we always assumed that the experimented commits contain accurate information of all co-changed functions. It is possible that developers made mistakes when applying multi-entity edits. Therefore, the imperfect evaluation data set based on developers' edits may influence our empirical comparison between \tool and ROSE. 
We share this limitation with prior work~\cite{Zimmermann:2004,Rolfsnes:2016, Islam:2018,Meng2013:lase,TanMing2015, Jiang2020, Wang2018CMSuggester}.
In the future, we plan to mitigate the problem by conducting user studies with developers. By carefully examining the edits made by developers and the co-changed functions recommended by tools, 
we can better assess the effectiveness of different tools. 
\section{Related Work}
\label{sec:relwork}
The related work includes empirical studies on JS code and related program changes, \red{JS bug detectors, and co-change recommendation systems.}

\subsection{Empirical Studies on JS Code and Related Program Changes}
\label{sec:empiricalRel}
Various studies were conducted to investigate JS code and related changes~\cite{Ocariza2013,Selakovic2016,Gao2017,Gyimesi2019,Silva2019}.
For instance, Ocariza et al.~conducted an empirical study of 317 bug reports from 12 bug repositories, to understand the root cause and consequence of each reported bug~\cite{Ocariza2013}. They observed that 65\% of JS bugs were caused by the faulty interactions between JS code and Document Object Models (DOMs).
Gao et al.~empirically investigated the benefits of leveraging static type systems (e.g., Facebook's Flow~\cite{flow} and Microsoft's TypeScript~\cite{typescript}) to check JS programs~\cite{Gao2017}. To do that, they manually added type annotations to buggy code and tested whether Flow and TypeScript reported an error on the buggy code. They observed that both Flow 0.30 and TypeScript 2.0 detected 15\% of errors, showing great potential of finding bugs. 

\red{Silva et a.~\cite{Silva2019} extracted changed source files from software version history, and revealed six co-change patterns by mapping frequently co-changed files to their file directories. Our research is different in three ways. First, we focused on software entities with finer granularities than files; we extracted the co-change patterns among classes, functions, variables, and statement blocks. Second, since unrelated entities are sometimes accidentally co-changed in program commits, we exploited the syntactic dependencies between entities to remove such data noise and to improve the quality of identified patterns. Third, \tool uses the identified patterns to further recommend co-changes with high quality. Wang et al.~\cite{yewang2018} recently conducted a study on multi-entity edits applied to Java programs, which study is closely relevant to our work. Wang et al.~focused on three kinds of software entities: classes, methods, and fields. They created CDGs for individual multi-entity edits, and revealed RCPs by comparing CDGs. The three most popular RCPs they found are: *CM$\xrightarrow{m}$CM (a callee method is co-changed with its caller(s)), *CM$\xrightarrow{m}$AM (a method is added, and one or more existing methods are changed to invoke the added method), and *CM$\xrightarrow{f}$AF (a field is added, and at least one existing method is changed to access the field).}

\red{Our research is inspired by Wang et al.'s work. We decided to conduct a similar study on JS programs mainly because JS is very different from Java. For instance, JS is weakly typed and has more flexible syntax rules; Java is strongly typed and variables must be declared before being used. JS is a script language and mainly used to make web pages more interactive; Java is used in more domains. We were curious whether developers' maintenance activities vary with the programming languages they use, and whether there are unique co-change patterns in JS programs. In our study, we adopted JS parsing tools, identified four kinds of entities in various ways, and did reveal some co-change patterns unique to JS programs because of the language's unique features. Surprisingly, the three most popular JS co-change patterns we observed match exactly with the Java co-change patterns mentioned above. Our study corroborates observations made by prior work. More importantly, it indicates that even though different programming languages provide distinct features, developers are likely to apply multi-entity edits in similar ways. This phenomenon sheds lights on future research directions of cross-language co-change recommendations.}


\subsection{JS Bug Detectors}
Researchers built tools to automatically detect bugs or malicious JS code~\cite{Cova2010,Ocariza2012,Schutt2012,Raychevnd2013,Amin2013,Park2014,Pradel2015,Pradel2018}.
For example,
EventRacer detects harmful data races in even-driven programs~\cite{Raychevnd2013}.
JSNose combines static and dynamic analysis to detect 13 JS smells in client-side code, where smells are code patterns that can adversely influence program comprehension and software maintenance~\cite{Amin2013}.
TypeDevil adopts dynamic analysis to warn developers about variables, properties, and functions that have inconsistent types~\cite{Pradel2015}. DeepBugs is a learning-based approach that formulates bug detection as a binary classification problem; it is able to detect accidentally swapped function arguments, incorrect binary operators, and incorrect operands in binary operations~\cite{Pradel2018}.
EarlyBird conducts dynamic analysis and adopts machine learning techniques for early identification of malicious behaviors of JavaScript code~\cite{Schutt2012}.

Some other researchers developed tools to suggest bug fixes or code refactorings~\cite{Feldthaus2011,MeaWad2012,Jensen2012,Ocariza2014,Monperrus2014,Selakovic2015,An2019}. With more details, Vejovis suggests program repairs for DOM-related JS bugs based on two common fix patterns: parameter replacements and DOM element validations~\cite{Ocariza2014}.
Monperrus and Maia built a JS debugger to help resolve ``crowd bugs'' (i.e., unexpected and incorrect outputs or behaviors resulting from the common and intuitive usage of APIs)~\cite{Monperrus2014}. Given a crowd bug, the debugger sends a code query to a server and retrieves all StackOverflow answers potentially related to the bug fix.
An and Tilevich built a JS refactoring tool to facilitate JS debugging and program repair~\cite{An2019}.
Given a distributed JS application, the tool first converts the program to a semantically equivalent centralized version by gluing together the client and server parts. After developers fixed bugs in the centralized version, the tool generates fixes for the original distributed version accordingly. In Model-Driven Engineering,
ReVision repairs incorrectly updated models by (1) extracting change patterns from version history, and (2) matching incorrect updates against those patterns to suggest repair operations~\cite{Ohrndorf2018}.

Our methodology is most relevant to the approach design of ReVision. However, our research is different in three aspects. First, our research focuses on entity-level co-change patterns in JS programs, while ReVision checks for consistencies different UML artifacts (e.g., the signature of a message in a sequence diagram must correspond to a method signature in the related class diagram). Second, the co-changed recommendation by \tool intends to complete an applied multi-entity edit, while the repair operations proposed by ReVision tries to complete consistency-preserving edit operations. Third, we conducted a large-scale empirical study to characterize multi-entity edits and experimented \tool with eight open-source projects, while ReVision was not empirically evaluated.

\subsection{Co-Change Recommendation Systems}
\label{sec:pre:mining}
Approaches were introduced to mine software version history and to extract co-change patterns~\cite{Silva2019,Gall1998,Gall2003,Shirabad2003,Zimmermann:2004,journals/tse/YingMNC04,Rolfsnes:2016, Islam:2018,Kagdi2007,Kagdi2012,Gethers2012,Zanjani:2014,Silva2015,Rolfsnes:2018}. Specifically,
Some researchers developed tools (e.g., ROSE) to mine the association rules between co-changed entities and to suggest possible changes accordingly~\cite{Zimmermann:2004,journals/tse/YingMNC04,Rolfsnes:2016, Islam:2018,Kagdi2007,Silva2015,Rolfsnes:2018}.
Some other researchers built hybrid approaches by combining
 information retrieval (IR) with association rule mining~\cite{Kagdi2012,Gethers2012,Zanjani:2014}.
Given a software entity $E$, these approaches use IR techniques to (1) extract terms from $E$ and any other entity and (2) rank those entities based on their term-usage overlap with $E$.
Meanwhile, these tools also apply association rule mining to commit history in order to rank entities based on the co-change frequency. In this way, if an entity $G$ has significant term-usage overlap with $E$ and has been co-changed a lot with $E$, then $G$ is recommended to be co-changed with $E$.

Shirabad et al.~developed a learning-based approach that predicts whether two given files should be changed together or not~\cite{Shirabad2003}. In particular, the researchers extracted features from software repository to represent the relationship between each pair of files, adopted those features of file pairs to train an ML model, and leveraged the model to predict whether any two files are relevant (i.e., should be co-changed) or not.
\tool is closely related to Shirabad et al.'s work. However, it is different in two aspects.
First, \tool predicts co-changed functions instead of co-changed files. With finer-granularity recommendations, \tool can help developers to better validate suggested changes and to edit code more easily. Second, our feature engineering for \tool is based on the quantitative analysis of frequent change patterns and qualitative analysis of the commonality between co-changed functions, while the feature engineering by Shirabad is mainly based on their intuitions. Consequently, most of our features are about the code commonality or co-evolution relationship between functions; while the features defined by Shirabad et al. mainly focus on file names/paths, routines referenced by each file, and the code comments together with problem reports related to each file.

\red{Wang et al.~built CMSuggester---an automatic approach to suggest method-level co-changes~\cite{Wang2018CMSuggester,Jiang2020}}. 
Different from CMSuggester, \tool is an ML-based instead of rule-based approach; it requires for data to train an ML model before suggesting changes while CMSuggester requires tool builders to hardcode the suggestion strategies. \tool recommends changes based on three RCPs: *CF$\xrightarrow{f}$CF, *CF$\xrightarrow{f}$AF, and *CF$\xrightarrow{v}$AV; CMSuggester recommends changes based on the last two patterns. \red{Our approach is more applicable. CMSuggester applies partial program analysis (PPA) to identify the referencer-referencee relationship between entities; it does not work when PPA is inapplicable. Meanwhile, \tool uses two alternative ways to infer the referencer-referencee relationship: typed-ast-util can accurately resolve bindings and link entities, while our heuristic approach links entities less accurately but is always applicable even if typed-ast-util does not work. Additionally,
our evaluation is more comprehensive. We evaluated \tool by integrating it with different ML algorithms and using the program data of eight open-source projects; CMSuggester was evaluated with the data of only four projects.}
 Overall, \tool is more flexible due to its usage of ML and is applicable to more types of co-change scenarios.

\section{Conclusion}
\label{sec:conclusion}

It is usually tedious and error-prone to develop and maintain JS code. To facilitate program comprehension and software debugging, we conducted an empirical study on multi-entity edits in JS projects and built an ML-based co-change recommendation tool \tool. 
Our empirical study explored the frequency and composition of multi-entity edits in JS programs, and investigated the syntactic and semantic relevance between frequently co-changed entities. In particular, we observed that 
(i) JS software developers frequently apply multi-entity edits while the co-changed entities are usually syntactically related; (ii) there are three most popular RCPs that commonly exist in all studied JS code repositories: *CF$\xrightarrow{f}$CF, *CF$\xrightarrow{f}$AF, and *CF$\xrightarrow{v}$AV; and (iii) among the entities matching these three RCPs, co-changed functions usually share certain commonality (e.g., common function invocations and common token subsequences).

Based on our study, we developed \tool, which tool extracts code features from the multi-entity edits that match any of the three RCPs, and trains an ML model with the extracted features to specially characterize relationship between co-changed functions. Given a new program commit or a set of entity changes that developers apply, the trained model extracts features from the program revision and recommends changes to complement applied edits as necessary. Our evaluation shows that \tool recommended changes with high accuracy and outperformed \red{two} existing techniques. In the future, we will investigate novel approaches to provide finer-grained code change suggestions and automate test case generation for suggested changes. 


\bibliography{ms}

\end{document}